\documentclass[a4paper,11pt]{article}
\pdfoutput=1

\usepackage{jcappub}

\usepackage[T1]{fontenc}

\usepackage{aas_macros}
\usepackage{bm}
\usepackage{multirow}

\title{\boldmath Multiscale analysis of the CMB temperature derivatives}

\author[a,b]{A.~Marcos-Caballero,}
\author[a]{E.~Mart\'\i nez-Gonz\'alez}
\author[a]{and P.~Vielva}

\affiliation[a]{Instituto de F\'isica de Cantabria, CSIC-Universidad de Cantabria,\\ Avda. de los Castros s/n, 39005 Santander, Spain.}
\affiliation[b]{Dpto. de F\'isica Moderna, Universidad de Cantabria,\\ Avda. los Castros s/n, 39005 Santander, Spain.}

\emailAdd{marcos@ifca.unican.es}
\emailAdd{martinez@ifca.unican.es}
\emailAdd{vielva@ifca.unican.es}

\abstract{We study the Planck CMB temperature at different scales through its
derivatives up to second order, which allows one to characterize the
local shape and isotropy of the field. The problem of having an
incomplete sky in the calculation and statistical characterization of
the derivatives is addressed in the paper. The analysis confirms the
existence of a low variance in the CMB at large scales, which is also
noticeable in the derivatives. Moreover, deviations from the standard
model in the gradient, curvature and the eccentricity tensor are
studied in terms of extreme values on the data. As it is expected, the
Cold Spot is detected as one of the most prominent peaks in terms of
curvature, but additionally, when the information of the temperature
and its Laplacian are combined, another feature with similar
probability at the scale of $10^\circ$ is also observed. However, the
$p$-value of these two deviations increase above the $6\%$ when they
are referred to the variance calculated from the theoretical fiducial
model, indicating that these deviations can be associated to the low
variance anomaly.  Finally, an estimator of the directional anisotropy
for spinorial quantities is introduced, which is applied to the
spinors derived from the field derivatives. An anisotropic direction
whose probability is $<1\%$ is detected in the eccentricity tensor.
}

\begin{document}
\maketitle
\flushbottom

\section{Introduction}

In the standard cosmological model, the high level of isotropy of the
Cosmic Microwave Background (CMB) observations are explained by
evoking a phase of exponential expansion in the early Universe, called
cosmic inflation. Following the standard predictions of inflation, the
initial perturbations are Gaussianly distributed in a homogeneous flat
space. Deviations in the isotropy and Gaussianity of the CMB
temperature field are important to constrain the particular model of
inflation, or even to explore new physics in the primordial Universe.

Although the recent measurements of the CMB have a good agreement with
the concordance model of cosmology \cite{planck132015}, there exists
evidences of particular deviations from the isotropy assumption as the
dipole modulation \cite{gordon2007,hoftuft2009}, parity violation
asymmetry \cite{land2005,kim2010,gruppuso2011} or the quadrupole and
octopole alignment \cite{copi2015}. Additionally, there are
indications that the CMB temperature presents a low variance
\cite{monteserin2008} and a lack of correlation
\cite{bennett2003,copi2015b} at large scales. Moreover, the Cold Spot
anomaly \cite{vielva2004,cruz2005} is characterized as an extreme on a
particular wavelet scale located in a specific region of the sky which
presents a deviation from Gaussianity. All these anomalies have the
common property that are especially dominated by the large scale
behaviour of the CMB.

The analysis of the temperature gradient and higher order derivatives
are useful for the characterization of the CMB anisotropies. For
instance, at small scales, the temperature gradient have been used to
reconstruct the matter density power spectrum using the gravitational
lensing effect over the CMB photons \cite{zaldarriaga1999}, as well as
to study the beam asymmetry systematic introduced by the scanning
strategy \cite{rathaus2014}. In a recent paper, the shape of the
large-scale peaks are analysed by considering the value of the
Laplacian and the eccentricity tensor at the centre of the peak
\cite{marcos2017}. In this work, we are interested in the large scale
behaviour of the CMB temperature derivatives, for which different
scales are analysed. If the scaling function is a Gaussian filter, the
multiscale analysis based on the scalar curvature (the Laplacian) is
equivalent to the Spherical Mexican Hat Wavelet
\cite{martinez2002,vielva2004}. On the other hand, the spinorial
derivatives (the gradient and the eccentricity tensor) have
information about the local directionality of the CMB temperature,
which have been studied previously by using the steerable wavelets
formalism \cite{wiaux2005,vielva2006,mcewen2015}. We present here a
joint analysis of the derivatives field for a wide range of scales
(from $1^\circ$ to $30^\circ$), paying attention to the extreme
deviations with respect to the standard model expectation.

The scalar curvature, as well as the temperature itself, allow one to
analyse the CMB in terms of rotational invariant quantities, which, in
particular, characterizes Cold Spot-like features on the sky. Besides
the scalar derivatives, the modulus of the gradient and the
eccentricity tensor can be studied in order to localize regions on the
sky with anomalous dipolar or quadrupolar local shape. Moreover, due
to the directional character of these derivatives, the gradient and
the eccentricity tensor can be also used to measure the isotropy of
the field. For this purpose, an estimator for spinorial quantities
based on the geodesic projection on a particular direction on the
sphere is introduced.

This paper is organised as follows: in Section~\ref{sec:theory}, the
calculation of the field derivatives in terms of the spherical
harmonic coefficients is introduced, whereas the data processing is
described in Section~\ref{sec:data}. A new formalism regarding the
statistics of the derivative fields in the presence of a mask is
introduced in Section~\ref{sec:pixel_cov}. The first analysis of this
work is performed in Section~\ref{sec:der_cov}, which consists in the
calculation and study of the covariance of the different derivatives
components. On the other hand, extreme values of the derivative fields
are analysed and compared with the standard model prediction in
Section~\ref{sec:extreme_der}. In addition, the directional analysis
of the spinorial derivatives is considered in
Section~\ref{sec:dir_analysis}, in order to quantify the isotropy of
the CMB temperature field. Finally, the conclusions of the paper are
exposed in Section~\ref{sec:conclusions}.

\section{Theoretical framework}
\label{sec:theory}

A field on the sphere is commonly described by the spherical
coordinates, and consequently, its derivatives are taken along the
directions determined by the local basis vectors $\mathbf{e}_\theta$
and $\mathbf{e}_\phi$. Additionally, it is useful to coordinate the
tangent plane in terms of the helicity basis $\mathbf{e}_\pm =
\mathbf{e}_\theta \pm i \mathbf{e}_\phi$, particularly when the
derivatives are expressed in the spherical harmonic space. The reason
for using this system of reference is that the covariant derivatives
in the helicity basis modify the spherical harmonics as the spin
raising/lowering operators, which simplifies the calculations. This
implies, in particular, that the derivatives of a field on the sphere
can be spanned in terms of the spin-weighted spherical harmonics,
whose spin depends on which type of derivative is considered. Since
spinors with different rank are statistically independent, we classify
the derivatives up to second order in two scalars ($s=0$), one vector
($s=1$) and one tensor ($s=2$), which allows to make an independent
analysis for each spin. The scalar derivatives are the temperature
field itself (zero-order derivative) and its Laplacian, which are
mutually correlated, specially at large scales. On the other hand, the
components of the gradient form a vector, while the local eccentricity
is a $2$-spin tensor determined by the second order
derivatives. Following the notation in \cite{marcos2016}, the
derivatives are given by
\begin{subequations}
\begin{equation}
\nu_R(\mathbf{n}) = \frac{1}{\sigma_\nu(R)} \sum_{\ell=0}^\infty
w_\ell(R) \
a_{\ell m} \ Y_{\ell m} (\mathbf{n}) \ ,
\label{eqn:nu}
\end{equation}
\begin{equation}
\kappa_R(\mathbf{n}) = \frac{1}{\sigma_\kappa(R)} \sum_{\ell=0}^\infty \frac{(\ell+1)!}{(\ell-1)!}
\ w_\ell(R) \ a_{\ell m} \ Y_{\ell m} (\mathbf{n}) \ ,
\end{equation}
\begin{equation}
\eta_R(\mathbf{n}) = \frac{1}{\sigma_\eta(R)} \sum_{\ell=1}^\infty
\sqrt{\frac{(\ell+1)!}{(\ell-1)!}} \ w_\ell(R) \ a_{\ell m} \ {}_{+1}Y_{\ell m}
(\mathbf{n}) \ ,
\end{equation}
\begin{equation}
\epsilon_R(\mathbf{n}) = \frac{1}{\sigma_\epsilon(R)} \sum_{\ell=2}^\infty
\sqrt{\frac{(\ell+2)!}{(\ell-2)!}} \ w_\ell(R) \ a_{\ell m} \ {}_{+2}Y_{\ell m}
(\mathbf{n}) \ ,
\label{eqn:eps}
\end{equation}
\label{eqn:derivatives}
\end{subequations}
where $w_\ell(R)$ represents the Fourier coefficients of the window
function corresponding to a particular angular scale $R$ (see
Section~\ref{sec:data} for the precise definition of this filter). In
the equations above, the scalar quantities $\nu$ and $\kappa$ are real
fields, while the spinors $\eta$ and $\epsilon$ are complex numbers,
since they have a directional character. The corresponding complex
conjugates are expressed in terms of the spherical harmonics with spin
$-1$ and $-2$, respectively. The derivatives in
eqs.~(\ref{eqn:derivatives}) are normalized by their corresponding
theoretical variances, which are calculated assuming a fiducial
model. They are expressed in terms of the angular power spectrum in
the following way:
\begin{subequations}
\begin{equation}
\sigma_\nu^2(R) = \sum_{\ell=0}^\infty \frac{2\ell+1}{4\pi}
w_\ell^2(R) \ C_\ell \ ,
\end{equation}
\begin{equation}
\sigma_\kappa^2(R) = \sum_{\ell=0}^\infty \frac{2\ell+1}{4\pi} \left[
  \frac{(\ell+1)!}{(\ell-1)!} \right]^2 w_\ell^2(R) \ C_\ell \ ,
\end{equation}
\begin{equation}
\sigma_\eta^2(R) = \sum_{\ell=1}^\infty \frac{2\ell+1}{4\pi}
\frac{(\ell+1)!}{(\ell-1)!} w_\ell^2(R) \ C_\ell \ ,
\end{equation}
\begin{equation}
\sigma_\epsilon^2(R) = \sum_{\ell=2}^\infty \frac{2\ell+1}{4\pi}
\frac{(\ell+2)!}{(\ell-2)!} w_\ell^2(R) \ C_\ell \ .
\end{equation}
\end{subequations}
The fact of normalizing the data by using a concrete theoretical model
does not introduces a bias in the analysis, since it can been seen as
a change of units in order to have unit variance quantities. The
theoretical fiducial model assumed throughout the paper is $\Omega_b
h^2 = 0.2222$, $\Omega_c h^2 = 0.1197$, $H_0 = 67.31 \ \mathrm{km/s}
\ \mathrm{Mpc^{-1}}$, $\tau = 0.078$, $n_s = 0.9655$ and $\ln(10^{10}
A_s) = 3.089$, which represent the Planck TT-lowP best-fit
cosmological parameters (\cite{planck132015}, table 3).

\section{Data processing}
\label{sec:data}

The CMB temperature data analysed in the paper correspond to the
cleaned maps delivered by the Planck collaboration. Since we are
studying large scale structures, the foreground contamination may be
important in the analysis and could introduce spurious signals. For
this reason, two of the four official temperature maps provided by
Planck are considered, namely, SEVEM and SMICA
\cite{planck092015}. These two maps are given in the Healpix
pixelation scheme with resolution $N_{\mathrm{side}} = 2048$
\cite{gorski2005}, and filtered by a Gaussian beam whose FWHM is
$5^\prime$ and the corresponding pixel window function. In terms of
the spherical harmonic coefficients, these maps have a band limit of
$\ell_\mathrm{max} = 4000$.

In order to consider CMB anisotropies at different scales, the maps
are filtered by a Gaussian function. The corresponding filter on the
sphere is obtained from the stereographic projection of the
two-dimensional Euclidean Gaussian distribution, whose Fourier
coefficients are given by:
\begin{equation}
w_\ell(R) = \exp \left[ -\frac{\ell\left( \ell + 1 \right)}{2R^2} \right] \ ,
\end{equation}
where the scale $R$, characterizing the width of the filter,
represents the standard deviation of a two-dimensional Gaussian
function in the Euclidean space. The linear scale $R$ is related to
the angular size $\theta$ on the sphere by $R = 2\tan
\frac{\theta}{2}$. However, the difference between these two
quantities is only important at large scales. Throughout the paper,
twenty angular scales from $1^\circ$ to $30^\circ$, which are chosen
with logarithmic steps, are considered in the analysis (see
table~\ref{tab:scales}).

\begin{table}
\begin{center}
\begin{tabular}{|c|c|c|c|} \hline
\emph{id}    & $R$ [deg] & $N_{\mathrm{side}}$ & $\ell_{\mathrm{max}}$ \\ \hline
  1 &  1.00 & 256 & 767 \\
  2 &  1.20 & 256 & 767 \\
  3 &  1.43 & 256 & 767 \\
  4 &  1.71 & 256 & 767 \\
  5 &  2.04 & 128 & 383 \\
  6 &  2.44 & 128 & 383 \\
  7 &  2.92 & 128 & 383 \\
  8 &  3.49 & 128 & 383 \\
  9 &  4.17 &  64 & 191 \\
 10 &  4.99 &  64 & 191 \\ \hline
\end{tabular}
\hspace{1cm}
\begin{tabular}{|c|c|c|c|} \hline
\emph{id}    & $R$ [deg] & $N_{\mathrm{side}}$ & $\ell_{\mathrm{max}}$ \\ \hline
 11 &  5.96 &  64 & 191 \\
 12 &  7.13 &  64 & 191 \\
 13 &  8.53 &  32 &  95 \\
 14 & 10.21 &  32 &  95 \\
 15 & 12.22 &  32 &  95 \\
 16 & 14.65 &  32 &  95 \\
 17 & 17.57 &  16 &  47 \\
 18 & 21.11 &  16 &  47 \\
 19 & 25.42 &  16 &  47 \\
 20 & 30.71 &  16 &  47 \\ \hline
\end{tabular}
\end{center}
\caption{Different scales considered in the paper. The first column
  indicates the labels used in Figure~\ref{fig:maxloc} to refer to
  that scale. The second column is the angular scale size $R$ measured
  in degrees. Finally, the third and fourth columns represent the
  resolution ($N_\mathrm{side}$) and the band limit
  ($\ell_{\mathrm{max}}$) used in the calculations of that particular
  scale, respectively.}
\label{tab:scales}
\end{table}

The data maps at the different scales are calculated by performing the
spherical harmonic transform of the temperature field up to the given
multipole and applying the filter $w_\ell(R)$ for each scale. Before
calculating the spherical harmonic coefficients, the maps are masked
with their respective confidence masks, and subsequently, the monopole
and dipole are removed in the remaining area. In this procedure, the
maps are deconvolved in order to remove the effective beam and pixel
window function present initially in the data. The derivative fields
are computed from the spherical harmonic coefficients following
eqs.~(\ref{eqn:nu}-\ref{eqn:eps}). The resulting maps are generated
again in the Healpix pixelisation scheme whose resolution depends on
the scale $R$ considered. This optimal resolution (in the sense of
working at the lowest resolution that retains all the useful
information of the filtered signal) is defined by taking into account
the properties of the stereographic projection:
\begin{equation}
N_{\mathrm{side}} \gtrsim \sqrt{\frac{1}{12}\left( 1 +
  \frac{4N}{R^2}\right)} \ ,
\label{eqn:nside}
\end{equation}
where $N$ represents the number of pixels in the area defined by a
circle of radius $R$. Since the values of $N_{\mathrm{side}}$ are only
powers of $2$, the Healpix resolution whose value is immediately
greater than the right-hand-side of eq.~(\ref{eqn:nside}) is
taken. The gradient, the curvature and the eccentricity tensor require
larger resolution than the temperature field, and, for this reason,
the value of $N = 56$ is chosen, which we have tested that provides
maps with the optimal resolution. Finally, the maps are generated
considering only multipoles up to $\ell_{\mathrm{max}} =
3N_{\mathrm{side}}-1$, once the corresponding pixel window function is
applied. Of course, the same procedure is applied to the simulations
used in the characterization of the statistical properties of the
data.

\section{Pixel covariance in the presence of a mask}
\label{sec:pixel_cov}

Since we are considering a field smoothed at different scales, the
mask applied to the data introduces different spurious correlations,
and reduces the variance in the region close to it for each convolved
version of the signal. In general, if the signal that we want to
analyse presents correlations, the particular geometry of the mask
becomes important and this effect is not trivial to consider
analytically. For this reason, the systematics introduced by the mask
are estimated by using a Monte Carlo methodology. Simulations of the
CMB anisotropies are generated and masked accordingly to the observed
sky in the data, and subsequently, the covariance of the derivatives
at different scales are calculated in each pixel. Since the maps are
smoothed after the mask is applied, the zeros imposed on unobserved
pixels affects to the unmasked region depending on the size of the
filter. For the largest scales considered in the paper, this effect
may be especially important.

We have developed a formalism in which the effect of the mask in each
pixel can be taken into account. The method is based on the
decomposition of the covariance between different masked fields at
each pixel as a linear transformation of the corresponding theoretical
covariance obtained in the full-sky limit:
\begin{equation}
\hat{\mathbf{C}}(p) = \mathbf{R}(p) \ \mathbf{C} \ \mathbf{R}^t(p) ,
\label{eqn:hatcov}
\end{equation}
where $\hat{\mathbf{C}}(p)$ is the covariance in the pixel $p$ for the
masked map, $\mathbf{C}$ is the full-sky covariance, which is
independent of the map location, and $\mathbf{R}(p)$ is the
transformation matrix relating them. As it is expected, this
transformation depends on the sky location due to the anisotropy
introduced by the mask. Since the mask geometry can be complicated,
the matrix $\mathbf{R}(p)$ is estimated by calculating simulations of
the particular masked fields under consideration.

The matrix which defines the linear transformation in
eq.~(\ref{eqn:hatcov}) is unique, imposing the condition that it is
lower triangular, in which case, it can be expressed as the product
$\mathbf{R}(p) = \hat{\mathbf{L}}(p) \mathbf{L}^{-1}$, where
$\hat{\mathbf{L}}(p)$ and $\mathbf{L}$ are the lower triangular
matrices obtained from the Cholesky decomposition of
$\hat{\mathbf{C}}(p)$ and $\mathbf{C}$, respectively. In the
particular case of two-dimensional covariances, the matrix
$\mathbf{R}(p)$ is explicitly given by
\begin{equation}
\mathbf{R}(p) = \left( \begin{array}{cc}
\frac{\hat{\sigma}_1}{\sigma_1} & 0 \\
\frac{\hat{\sigma}_2}{\sigma_1} \left( \hat{\rho} - \rho
\sqrt{\frac{1-\hat{\rho}^2}{1-\rho^2}} \right) & 
\frac{\hat{\sigma}_2}{\sigma_2} \sqrt{\frac{1-\hat{\rho}^2}{1-\rho^2}}
\end{array} \right) \ ,
\label{eqn:rmatrix}
\end{equation}
where $\sigma_1$ and $\sigma_2$ are the r.m.s. of the two considered
random variables, and $\rho$ is the correlation coefficient between
them. The same quantities, but with the hat notation, indicates the
corresponding variables when the field is masked. Notice that
$\hat{\sigma}_1$, $\hat{\sigma}_2$ and $\hat{\rho}$ depend on the sky
location due to the anisotropy introduced by the mask. Since the
components of the matrix $\mathbf{R}(p)$ are formed by ratios of
masked and unmasked quantities, it converges faster than
$\hat{\mathbf{C}}(p)$ when they are estimated with simulations. For
instance, for pixels away from the mask, where the effect of the
smoothing on the derivatives is negligible, the matrix $\mathbf{R}$
approaches to the identity with practically zero variance. On the
other hand, if the covariance $\hat{\mathbf{C}}(p)$ is calculated
directly, it is needed more simulations to converge with a desired
precision, even in points for which the mask has no effect. For this
reason, the covariance of the the masked fields $\hat{\mathbf{C}}(p)$
is calculated from eq.~(\ref{eqn:hatcov}), where the transformation
matrix $\mathbf{R}(p)$ is estimated with simulations and the full-sky
covariance $\mathbf{C}$ is computed theoretically.

Besides for the calculation of $\hat{\mathbf{C}}(p)$, the matrix
$\mathbf{R}(p)$ in eq.~(\ref{eqn:rmatrix}) is used to construct an
estimator of the one-point covariance of the derivative fields (see
Section~\ref{sec:der_cov}), and, additionally, to take into account
the observed low variance of the CMB \cite{monteserin2008} in the
analysis performed in Sections~\ref{sec:extreme_der} and
\ref{sec:dir_analysis}.

\section{Covariance of the derivatives}
\label{sec:der_cov}

One of the anomalies in the CMB data is the low variance of the
temperature field at large scales, which have been confirmed by
several analyses
\cite{monteserin2008,cruz2011,gruppuso2013,planck162015}. In this
work, these studies are extended to the variance of each derivative
field and the correlation between $\nu$ and $\kappa$. In order to take
into account the smoothing effects on a incomplete sky, it is proposed
an estimator that considers the different variance at each pixel. In
particular, for two correlated variables given by the vector
$\mathbf{x}(p) = (x_1(p),x_2(p))$, the estimator of their covariance
matrix is
\[
\mathbf{S}^2_{\mathbf{x}} = \frac{1}{N_{\mathrm{pix}}} \sum_{p}
\mathbf{R}^{-1}(p) \mathbf{x}(p) \left[ \mathbf{R}^{-1}(p)
  \mathbf{x}(p) \right]^t =
\]
\begin{equation}
= \frac{1}{N_{\mathrm{pix}}} \sum_{p}
\mathbf{R}^{-1}(p) \left( \begin{array}{cc} x_1^2(p) & x_1(p) x_2(p)
  \\ x_1(p) x_2(p) & x_2^2(p) \\
\end{array} \right) \left[ \mathbf{R}^{-1}(p) \right]^t \ ,
\label{eqn:covariance_est}
\end{equation}
where the anisotropy of the field due to the mask is considered by
using the matrix $\mathbf{R}(p)$ defined in eq.~(\ref{eqn:rmatrix})
and $N_{\mathrm{pix}}$ represents to number of observed pixels. This
expression is obtained by inverting eq.~(\ref{eqn:hatcov}) to express
the full-sky covariance as a function of the covariance from the data,
which is estimated as the product of the two fields at each pixel, and
averaging over the observed sky, once the correction given by the
inverse of the matrix $\mathbf{R}(p)$ is applied. From
eq.~(\ref{eqn:covariance_est}), it is possible to see that the
estimator of the variance of the field $x$ is
\begin{equation}
S^2_x = \frac{\sigma^2_x}{N_{\mathrm{pix}}} \sum_{p}
\frac{|x(p)|^2}{\hat{\sigma}^2_x(p)} \ ,
\label{eqn:variance_est}
\end{equation}
where we have included the modulus of $x$ in order to generalize this
expression for complex fields, like the gradient and the eccentricity
tensor. These variances are calculated as a sum over all observed
pixels weighted by the ratio $\sigma^2_x/\hat{\sigma}^2_x(p)$, which
represents the anisotropy introduced by the mask. In the case of
pixels away from the mask, the weights approach to unity, recovering
the standard variance estimator for full sky maps. Averaging
eq.~(\ref{eqn:variance_est}), the variance of the field $x$ is
obtained, which implies that the estimator $S^2_x$ is unbiased.

In addition to the variances, the cross-correlation between the
variables $x$ and $y$ can be calculated using the following estimator,
which is obtained from the off-diagonal component of
eq.~(\ref{eqn:covariance_est}):
\begin{equation}
S_{xy} = \frac{\sigma_x\sigma_y}{N_{\mathrm{pix}}}
\sum_{p}
\frac{x(p)y(p)}{\hat{\sigma}_x(p)\hat{\sigma}_y(p)}
\sqrt{\frac{1-\rho^2}{1-\hat{\rho}^2(p)}} +
\frac{\sigma_x\sigma_y}{N_{\mathrm{pix}}} \sum_{p} \frac{
  x^2(p)}{\hat{\sigma}^2_x(p)} \left( \rho - \hat{\rho}(p)
\sqrt{\frac{1-\rho^2}{1-\hat{\rho}^2(p)}} \right) \ .
\end{equation}
The fact that the mask modifies the correlation coefficient of the two
variables introduces a second term on the right-hand-side of this
equation in order to prevent a biased estimation of the
cross-correlation. In the particular case of having full-sky maps,
$S_{xy}$ coincides with the standard estimator of the
cross-correlation of two variables.

\begin{figure}
\begin{center}
\includegraphics[scale=0.39]{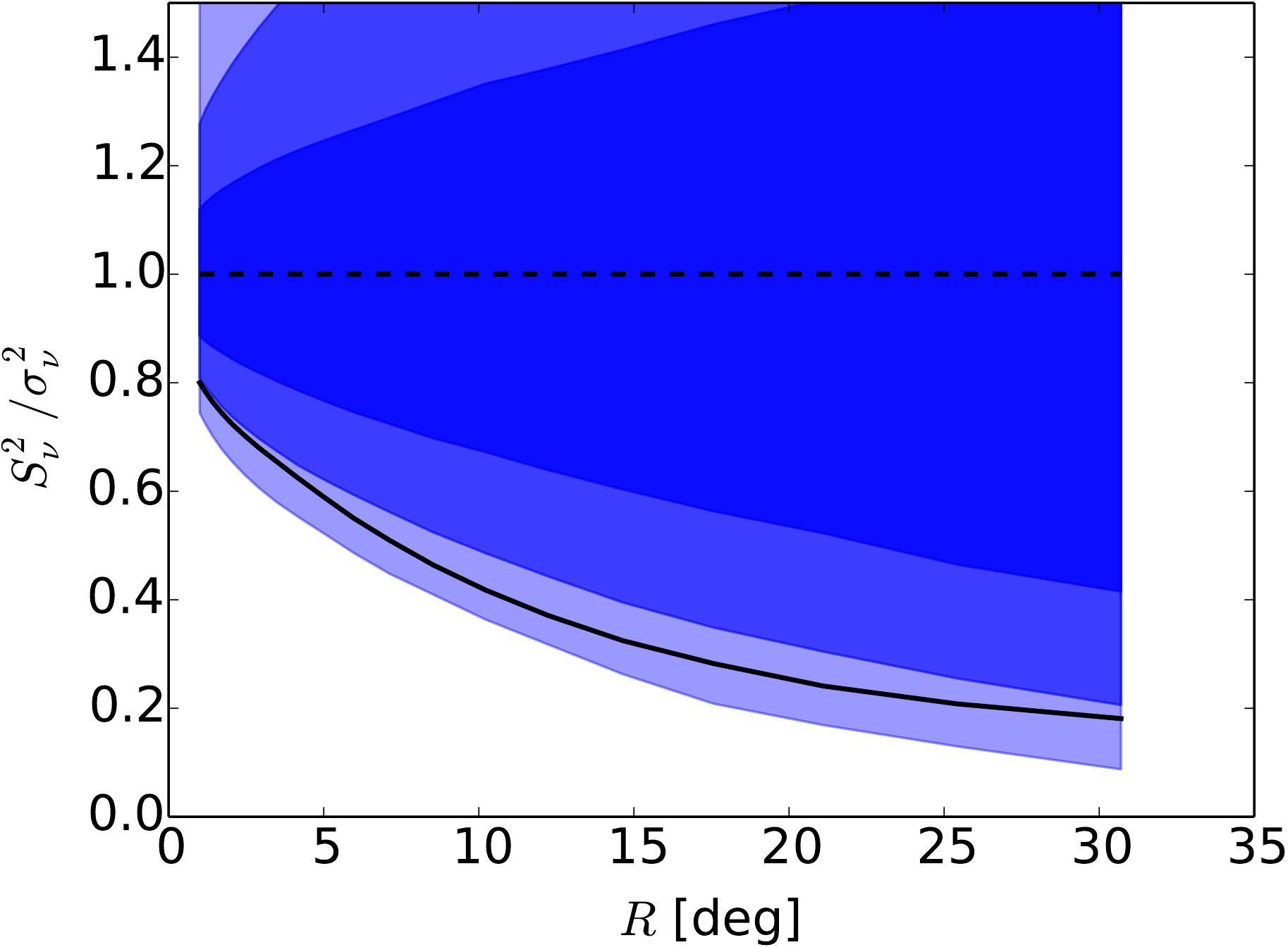}
\includegraphics[scale=0.39]{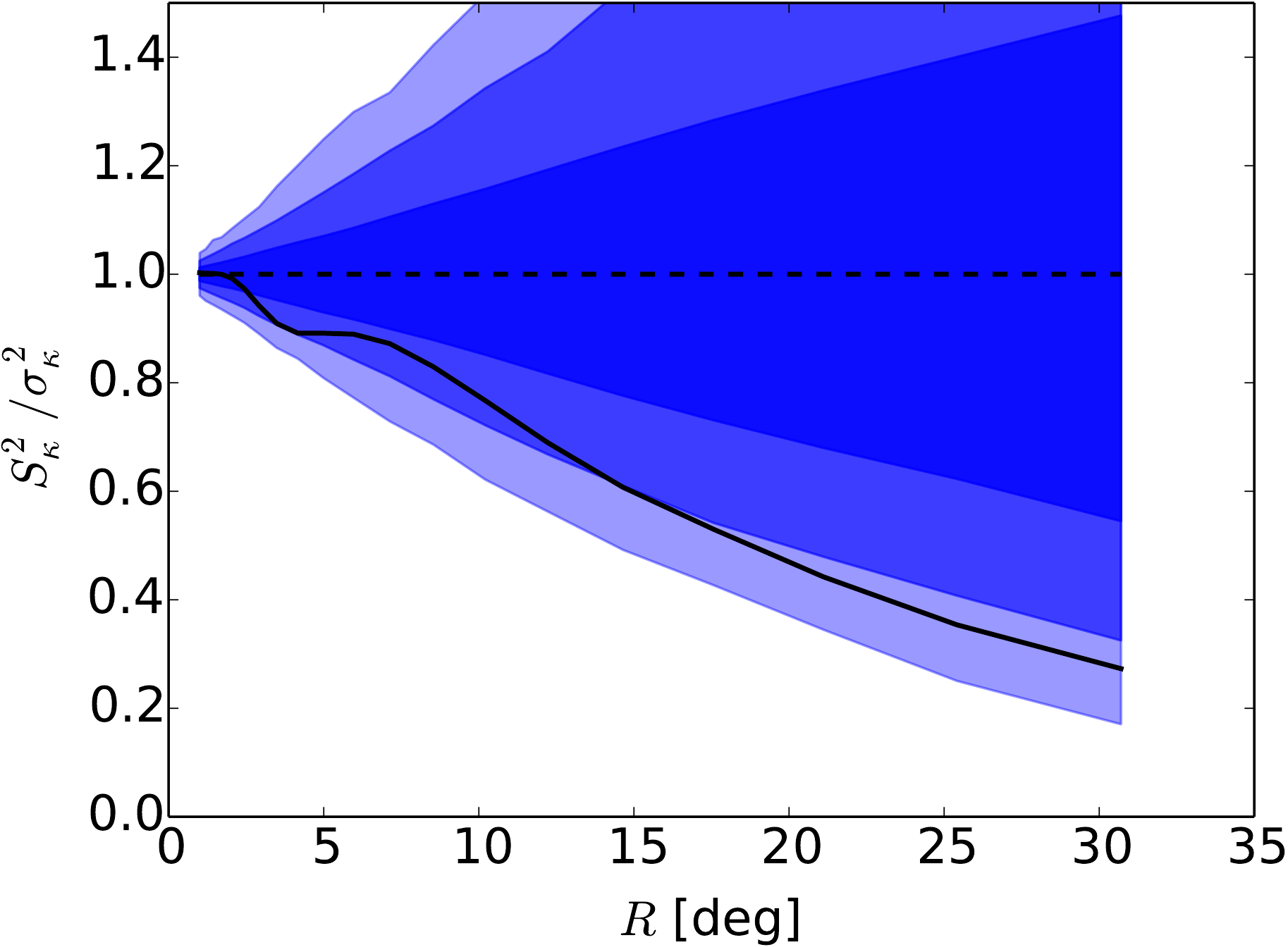}
\includegraphics[scale=0.39]{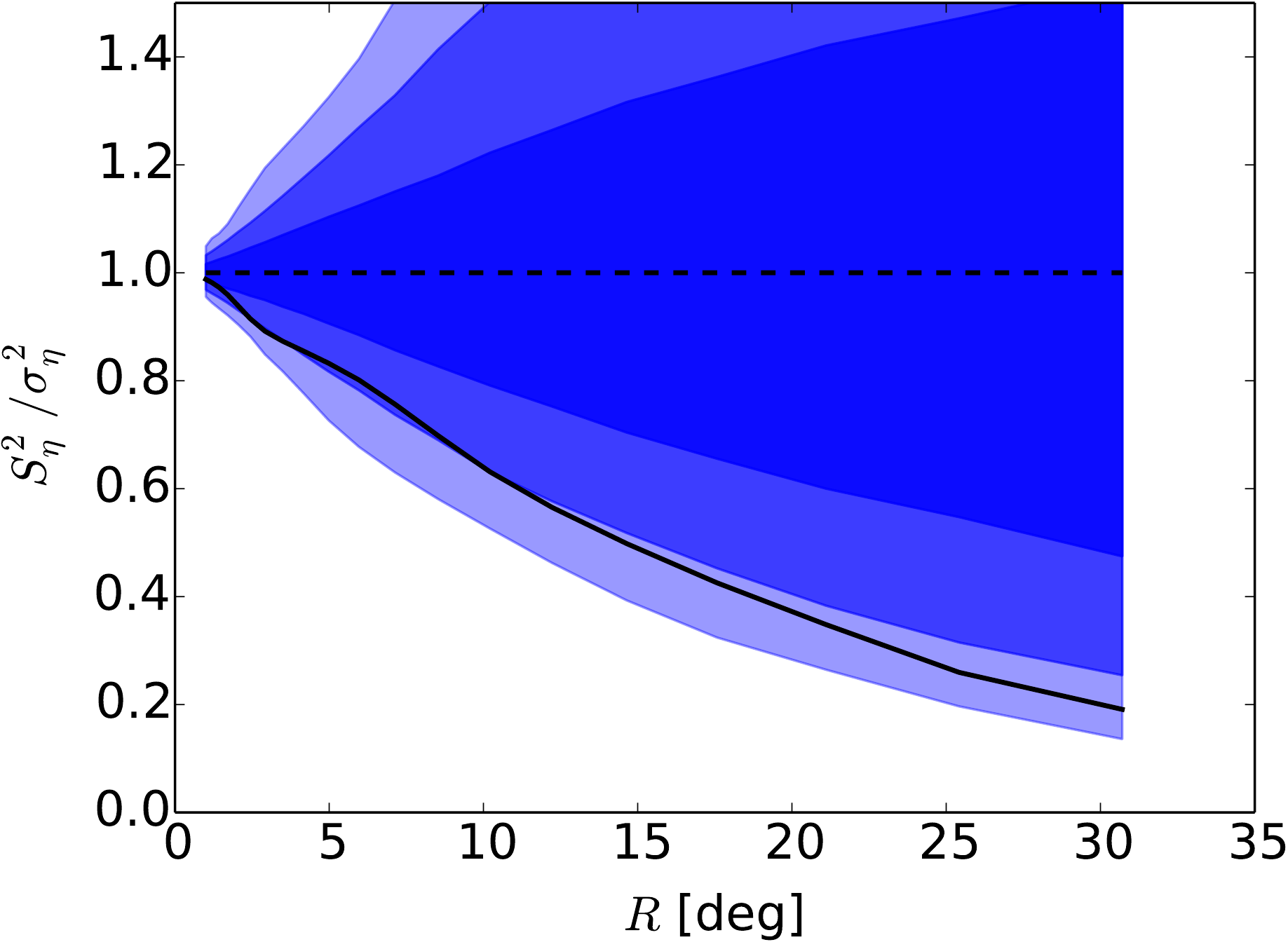}
\includegraphics[scale=0.39]{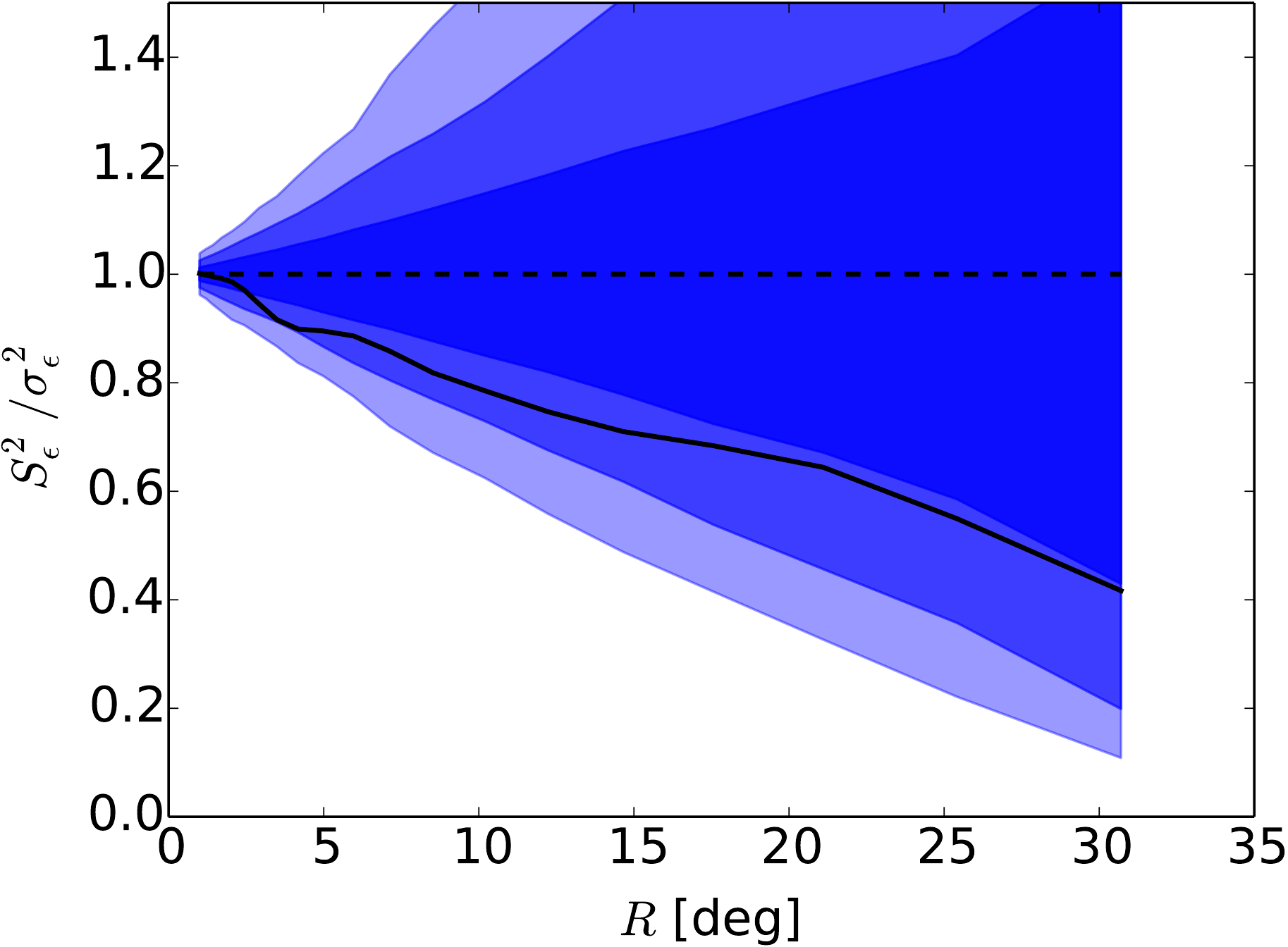}
\end{center}
\caption{Variances of the derivatives fields $\nu$, $\kappa$, $\eta$
  and $\epsilon$ as a function of the scale $R$, for the Planck SEVEM
  map. The three contours correspond to the one, two and three sigma
  regions.}
\label{fig:variances}
\end{figure}

\begin{figure}
\begin{center}
\includegraphics[scale=0.39]{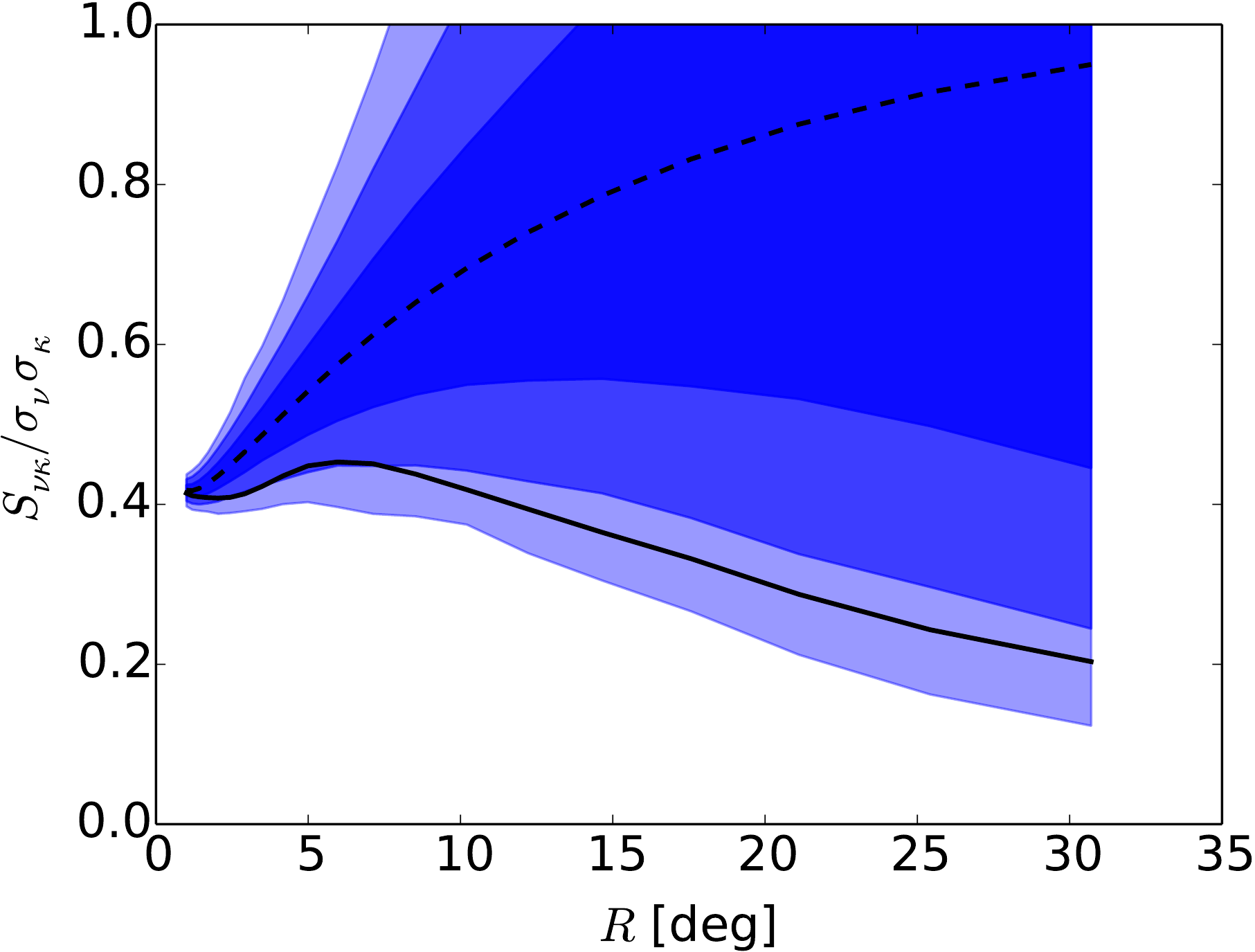}
\includegraphics[scale=0.39]{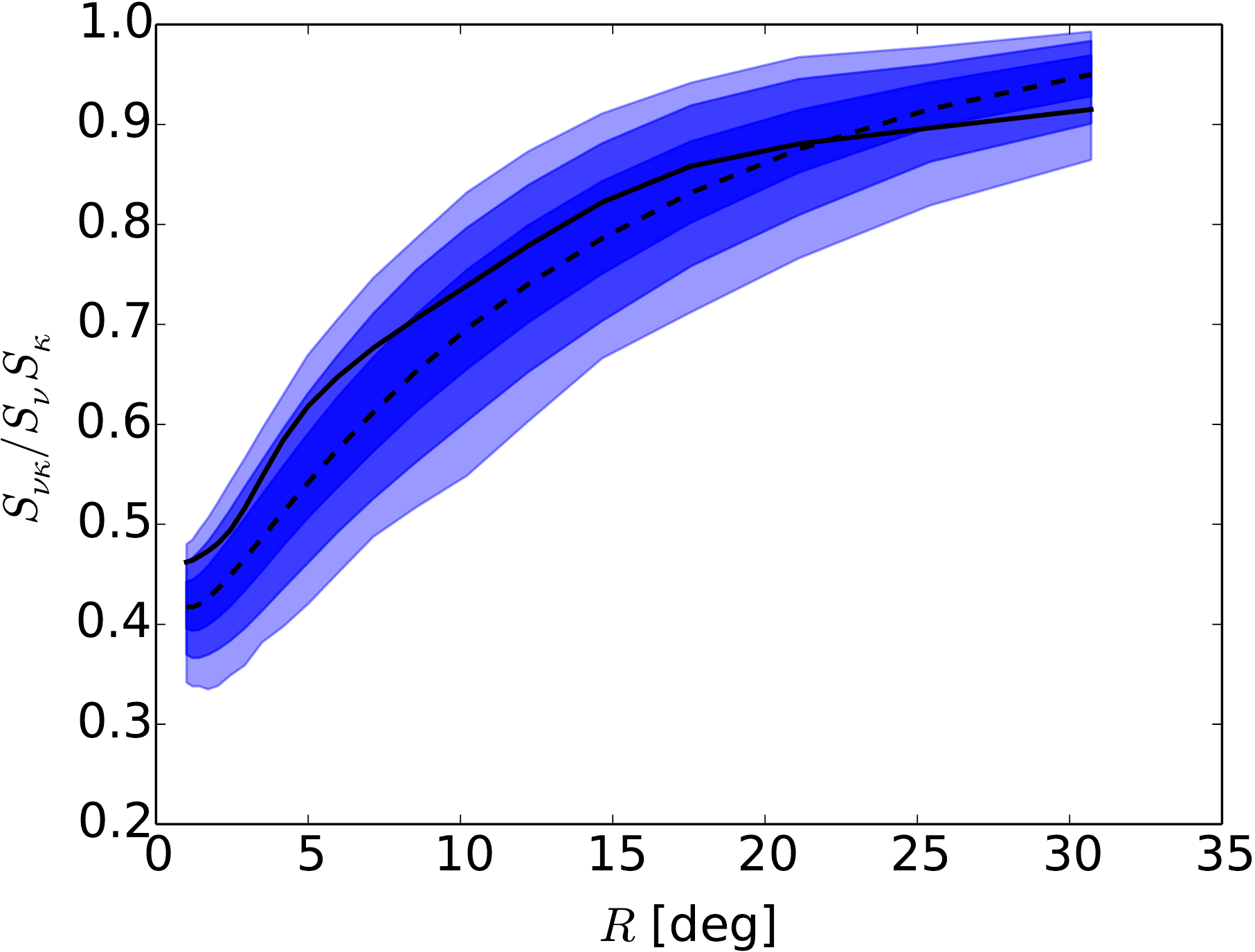}
\end{center}
\caption{Cross-correlation $S_{\nu\kappa}$ between the temperature and
  the curvature (left), and its corresponding correlation coefficient
  (right) as a function of the scale, obtained from the Planck SEVEM
  map. The contours correspond to the one, two and three sigma
  regions.}
\label{fig:cross_rho_nak}
\end{figure}

In figure~\ref{fig:variances}, it is represented the variance of the
derivatives fields as a function of the scale. The variance of the
normalized temperature field $\nu$ presents a low variance with
significance greater than 2-$\sigma$ for scales $> 2^\circ$. This
result is in agreement with previous analyses of the low variance of
the CMB temperature at large scales \cite{planck162015}. Regarding
the derivatives, the gradient $\eta$ and the curvature $\kappa$ also
have a low variance at large scale, which are significant for $R >
15^\circ$, whereas the eccentricity remains below the 2-$\sigma$ level
for all scales. Additionally, the low variance is also manifested in
the cross-correlation of $\nu$ and $\kappa$ for scales $R > 5^\circ$,
but, on the other hand, this anomaly is no present in the
cross-correlation coefficient, which is calculated normalizing by the
observed standard deviations instead of their theoretical values
(figure~\ref{fig:cross_rho_nak}). This is an indication that, despite
of low variance in $\nu$ and $\kappa$, the correlation of these two
variables is compatible with the theoretical expectation. Both CMB
Planck maps, SEVEM and SMICA, give similar results in terms of the
variance of the derivative fields.

\begin{figure}
\begin{center}
\includegraphics[scale=0.39]{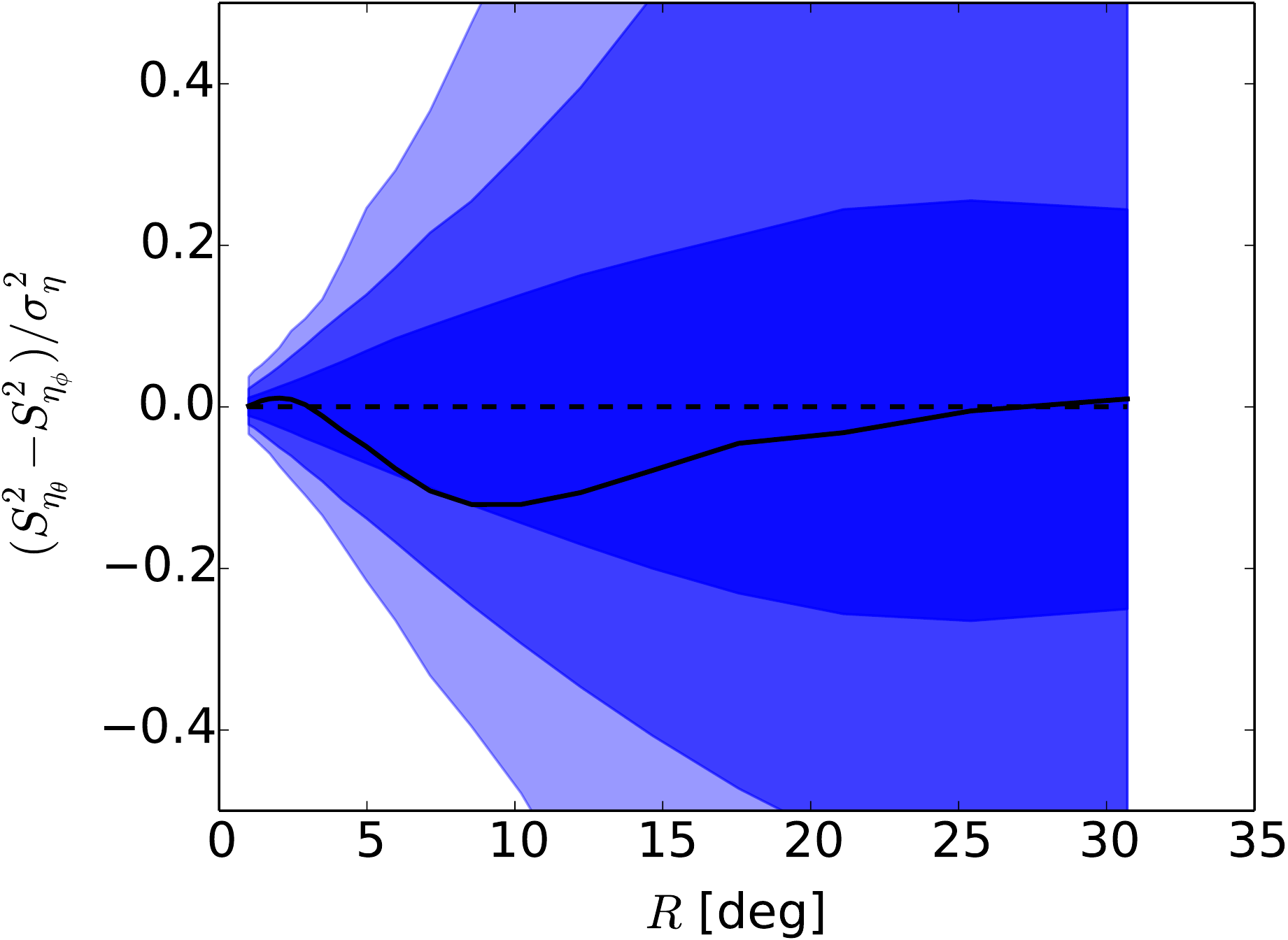}
\includegraphics[scale=0.39]{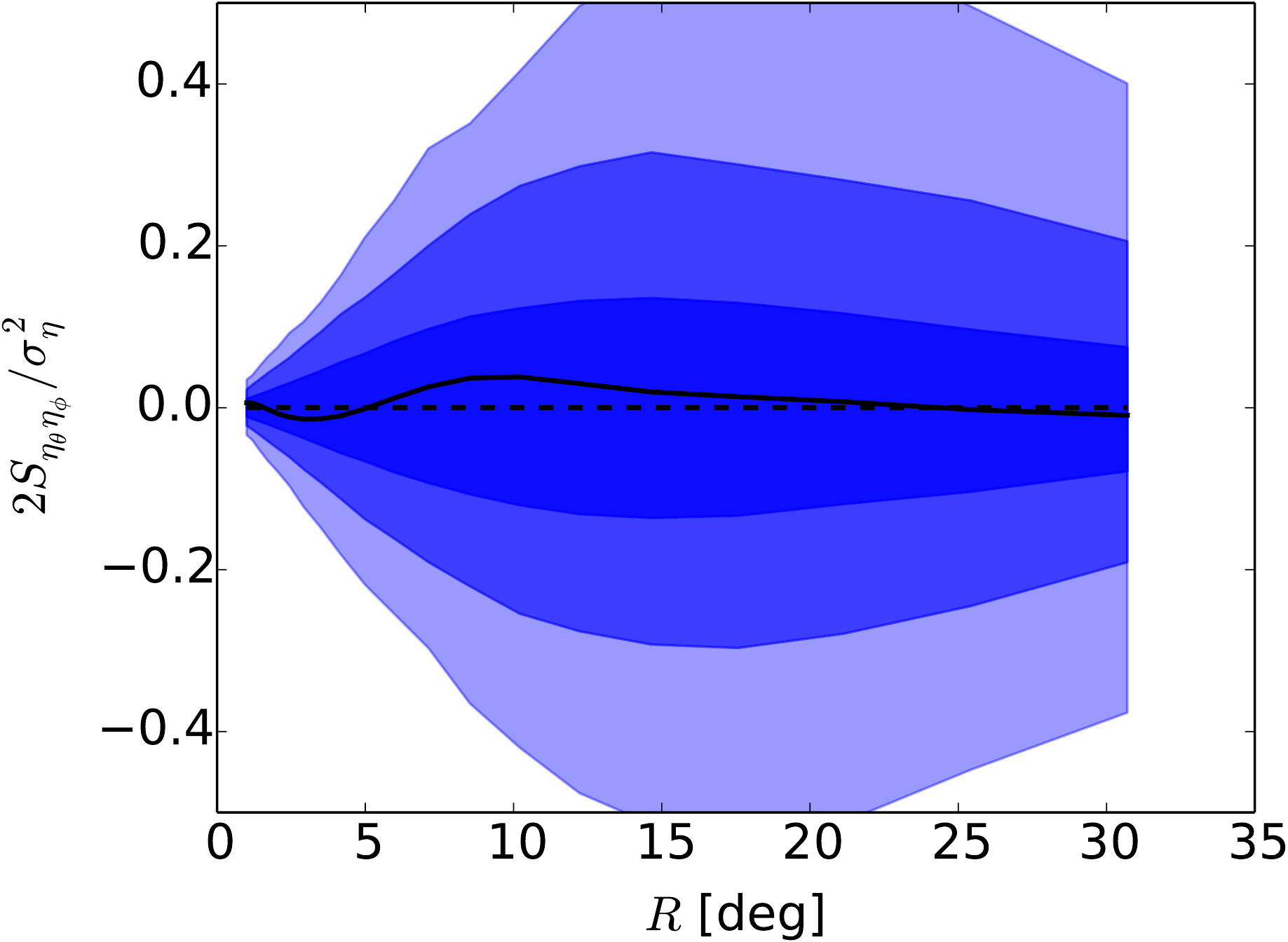}
\includegraphics[scale=0.39]{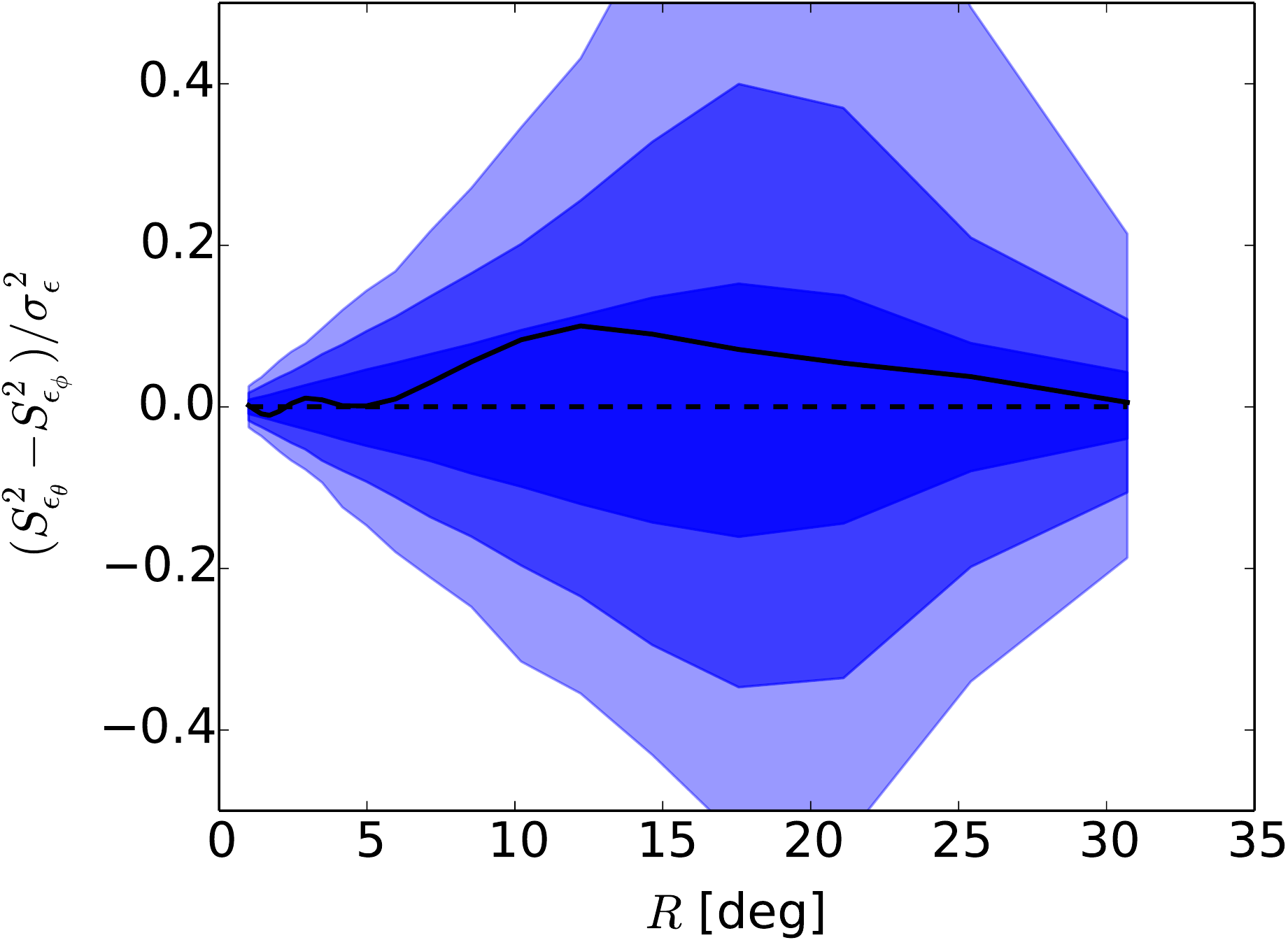}
\includegraphics[scale=0.39]{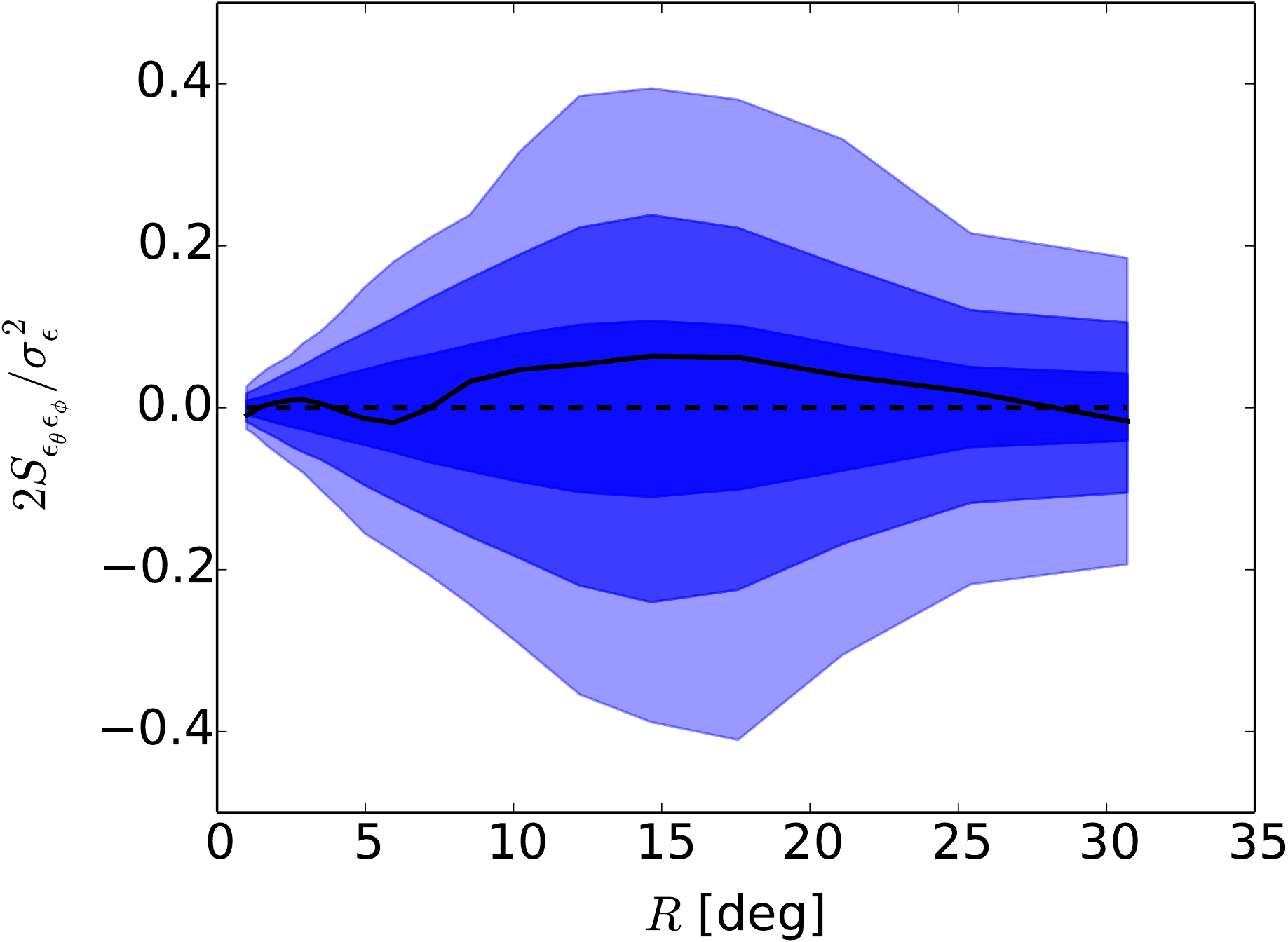}
\end{center}
\caption{Difference of the variances (left) and cross-correlation
  (right) of the two component of the spinors $\eta$ (upper row) and
  $\epsilon$ (bottom row). The curves are obtained from the Planck
  SEVEM data. The contours correspond to the one, two and three sigma
  regions.}
\label{fig:diff_cross_eta_eps}
\end{figure}

In addition to the analysis of the variance, it is possible to study
the local isotropy of the field looking at the variance of the
different components of the spinorial derivatives. Since the
cosmological principle implies that any statistical quantity does not
depend on the direction considered, the variance of each component
must be the same as well as the correlation between them must
vanish. In figure~\ref{fig:diff_cross_eta_eps}, the difference of the
variances and the cross-correlation between the two components of
$\eta$ and $\epsilon$ are depicted, showing that the data are
compatible with the isotropy of the field. Notice that these
quantities depend on the particular choice of the coordinate system,
indicating in this case, an alignment with the galactic north-south
direction, which corresponds to the $z$-axis of the spherical
coordinates. A more general analysis of the statistical isotropy of
the derivatives, which is independent of the particular coordinate
system assumed, is performed in Section~\ref{sec:dir_analysis}.

\section{Extreme deviations in the derivatives fields}
\label{sec:extreme_der}

In order study deviations of the CMB temperature field and its
derivatives from the standard model prediction, the tail probability
of the one-point distribution for each pixel is calculated. Assuming
the predictions of the simplest models of inflation, the CMB
fluctuations are Gaussianly distributed and, therefore, the variables
$\nu$, $\kappa$ and the components of the spinors $\eta$ and
$\epsilon$ are also Gaussian variables, since they are calculated
applying linear operators over the CMB temperature field.

For the scalar degrees of freedom $\nu$ and $\kappa$, we define the
following quantities:
\begin{subequations}
\begin{equation}
\chi^2_\nu(p) = \frac{\nu^2(p)}{\hat{\sigma}^2_\nu(p)} \sigma^2_\nu
\ ,
\label{eqn:chi2_nu}
\end{equation}
\begin{equation}
\chi^2_\kappa(p) = \frac{\kappa^2(p)}{\hat{\sigma}^2_\kappa(p)}
\sigma^2_\kappa \ ,
\end{equation}
\begin{equation}
\chi^2_s(p) = \mathbf{s}(p) \hat{\mathbf{C}}_{s}^{-1}(p) \mathbf{s}(p)
\ ,
\end{equation}
\label{eqn:chi2_scal}
\end{subequations}
where the vector $\mathbf{s} = (\nu,\kappa)$ is composed by the
temperature and curvature fields, and $\hat{\mathbf{C}}_s(p)$ denotes
the covariance of $\mathbf{s}$ in the pixel $p$, which is calculated
from eq.~(\ref{eqn:hatcov}). If the temperature field is Gaussian,
these quantities are distributed according to the $\chi^2$ probability
density. Whilst the $\chi^2$ distribution has one degree of freedom in
the case of $\chi^2_\nu$ and $\chi^2_\kappa$, the combination
$\chi^2_s$ has two degrees of freedom. In a similar way, it is
computed the $\chi^2$ quantities for the gradient and the eccentricity
tensor:
\begin{subequations}
\begin{equation}
\chi^2_\eta(p) = {\bm \eta}(p) \hat{\mathbf{C}}_{\eta}^{-1}(p) {\bm
  \eta}(p) \ ,
\end{equation}
\begin{equation}
\chi^2_\epsilon(p) = {\bm \epsilon}(p)
\hat{\mathbf{C}}_{\epsilon}^{-1}(p) {\bm \epsilon}(p) \ ,
\label{eqn:chi2_eps}
\end{equation}
\label{eqn:chi2_spin}
\end{subequations}
where, in this case, the vectors ${\bm \eta}$ and ${\bm \epsilon}$
denote the components of the spinors $\eta$ and $\epsilon$,
respectively.

In order to quantify deviations from the standard model, we compute
the logarithm of the tail probability of $\chi^2$ variables defined in
eqs.~(\ref{eqn:chi2_scal}-\ref{eqn:chi2_spin}):
\begin{equation}
r_x(p) = - \ln \mathrm{P}\{ \chi^2 > \chi^2_x(p) \} \ ,
\end{equation}
where $x$ represents the field considered ($\nu$, $\kappa$, $s$,
$\eta$ or $\epsilon$). One advantage of this quantity is that it is
distributed following the exponential probability density,
independently of the degrees of freedom of the $\chi^2$ variable
considered (the logarithm of a variable that is uniformly distributed
has an exponential probability density.). Possible anomalies on the
derivative fields are identified by looking at large values of
$r_x(p)$, which correspond to large deviations of $\chi^2_x(p)$. In
particular, the maximum of $r_x(p)$ can be computed:
\begin{equation}
r_x = \max_p \{ r_x(p) \} \ .
\end{equation}
Due to the intrinsic correlations on the field, the values of $r_x(p)$
for different pixels are correlated, which modifies the distribution
of the maximum $r_x$, particularly at large scales, where the
correlations dominates the field. For this reason, the probability
distribution of $r_x$ is calculated using a Monte Carlo method with
$5000$ simulations.

\begin{figure}
\begin{center}
\includegraphics[scale=0.39]{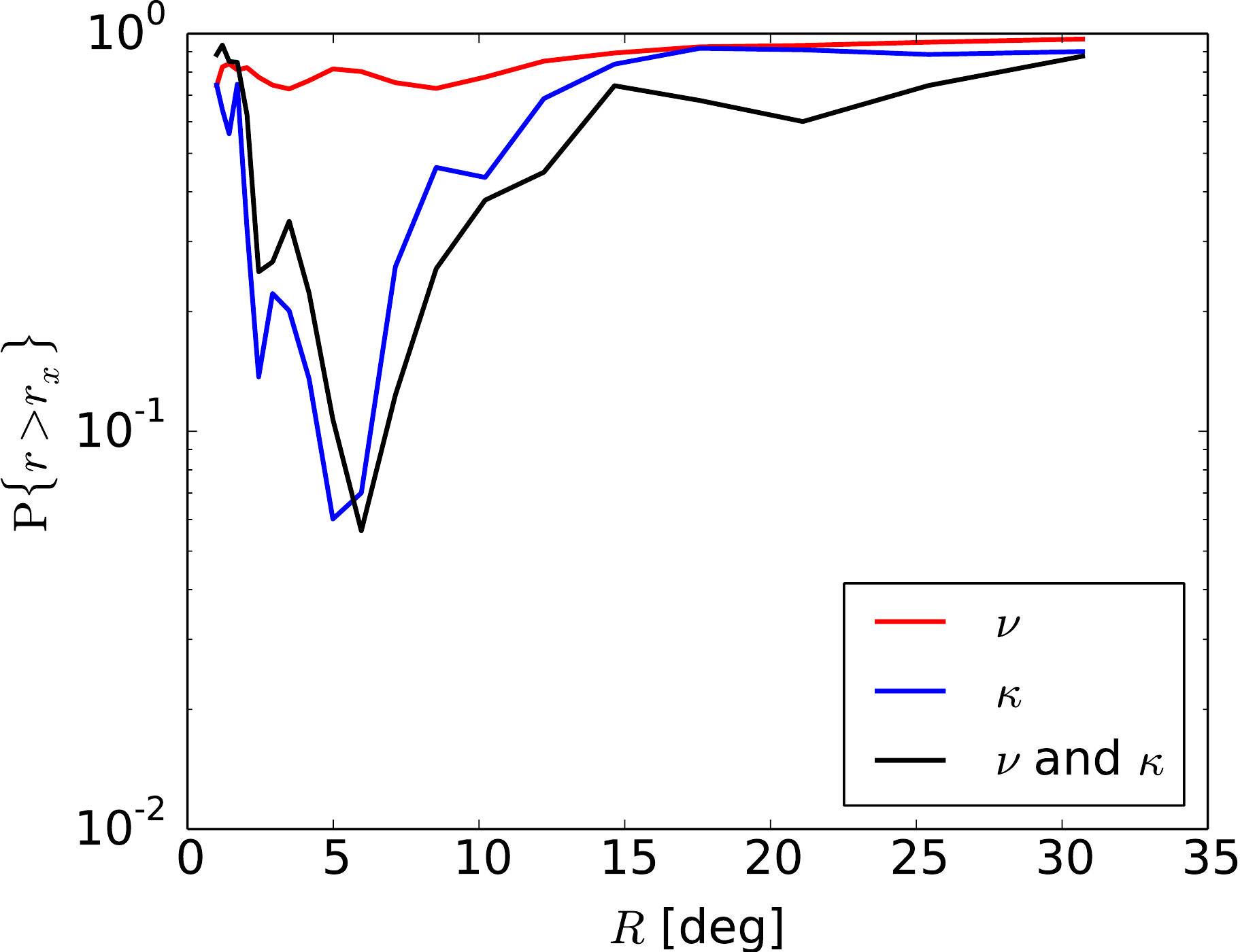}
\includegraphics[scale=0.39]{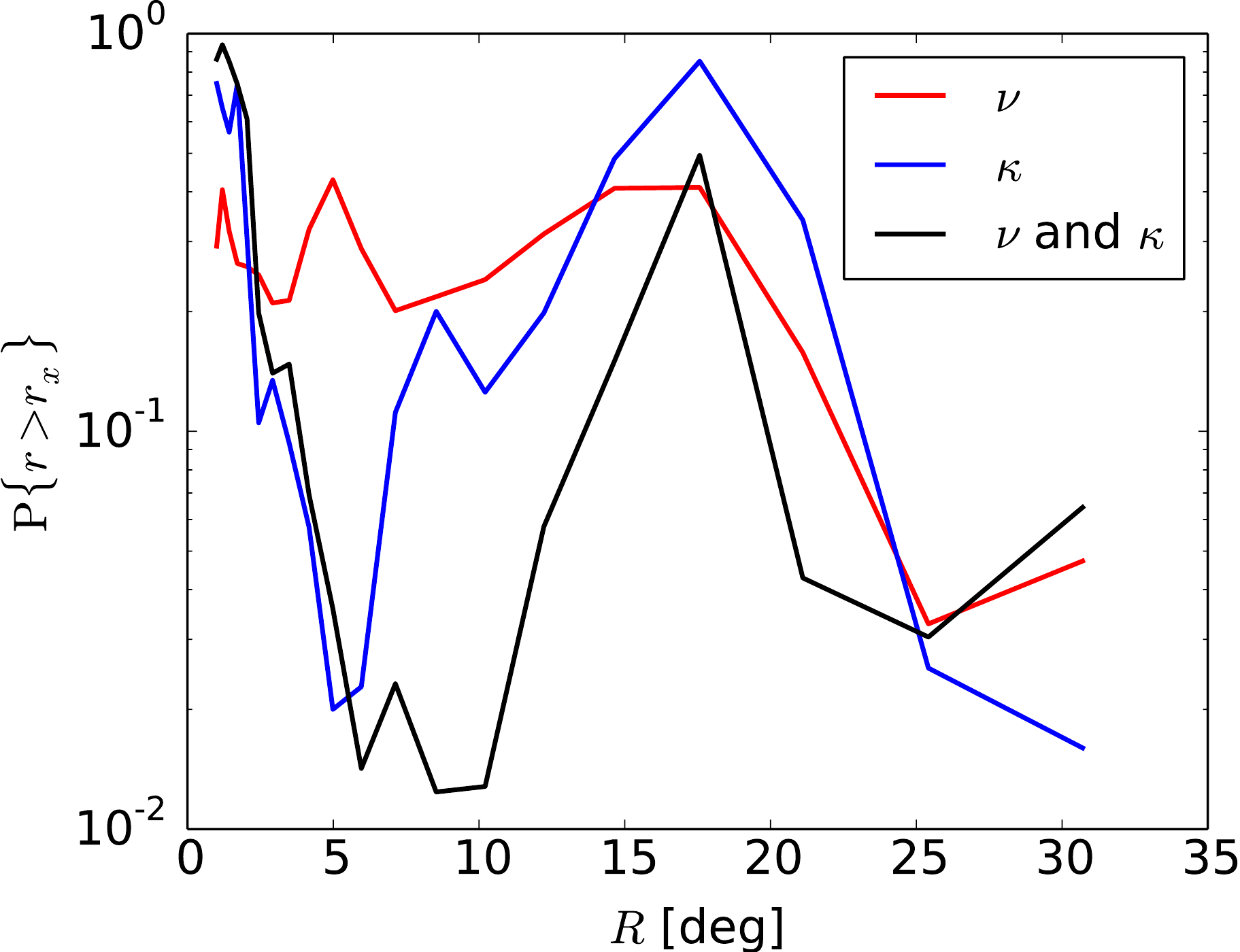}
\includegraphics[scale=0.39]{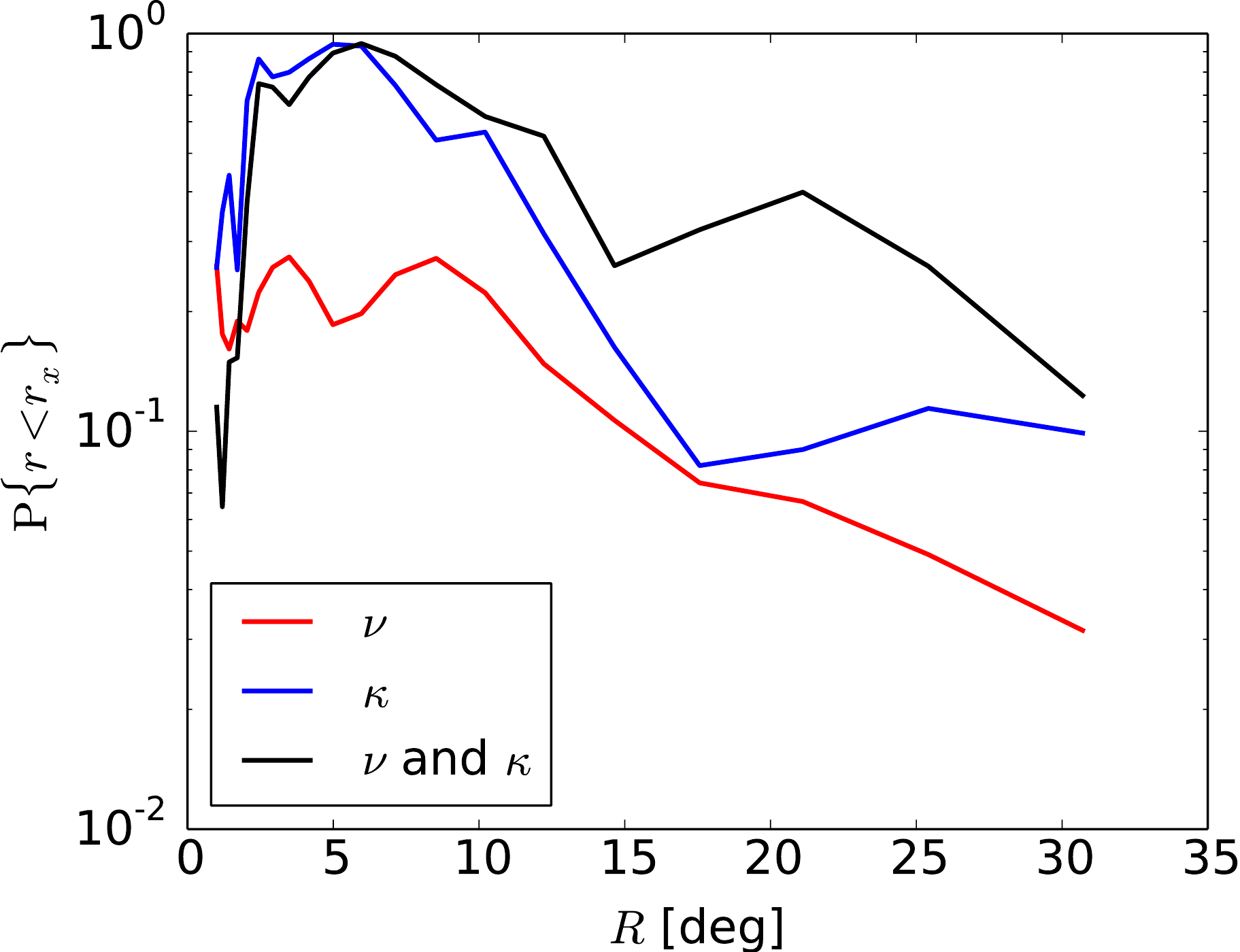}
\includegraphics[scale=0.39]{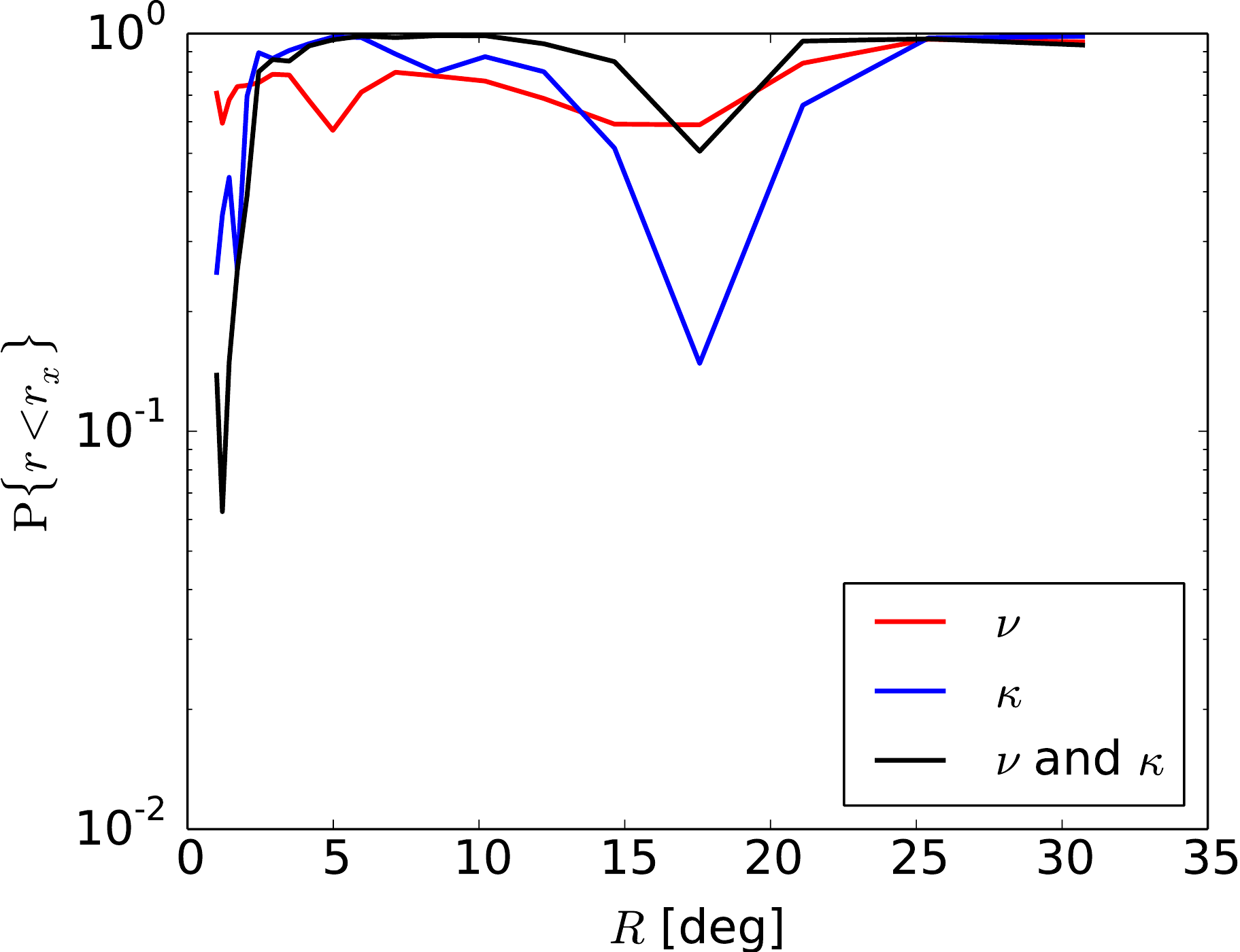}
\end{center}
\caption{Upper tail probability (upper row) and lower tail probability
  (bottom row) of the maximum value $r_x$ for the temperature, the
  curvature and the combination of both, obtained from the Planck
  SEVEM map. In the first column, the covariance is calculated from
  the theoretical fiducial model, whereas the probabilities in the
  second column are calculated using the covariance obtained from the
  data, in order to take into account the low variance.}
\label{fig:pvalue_r_nak}
\end{figure}

\begin{figure}
\begin{center}
\includegraphics[scale=0.39]{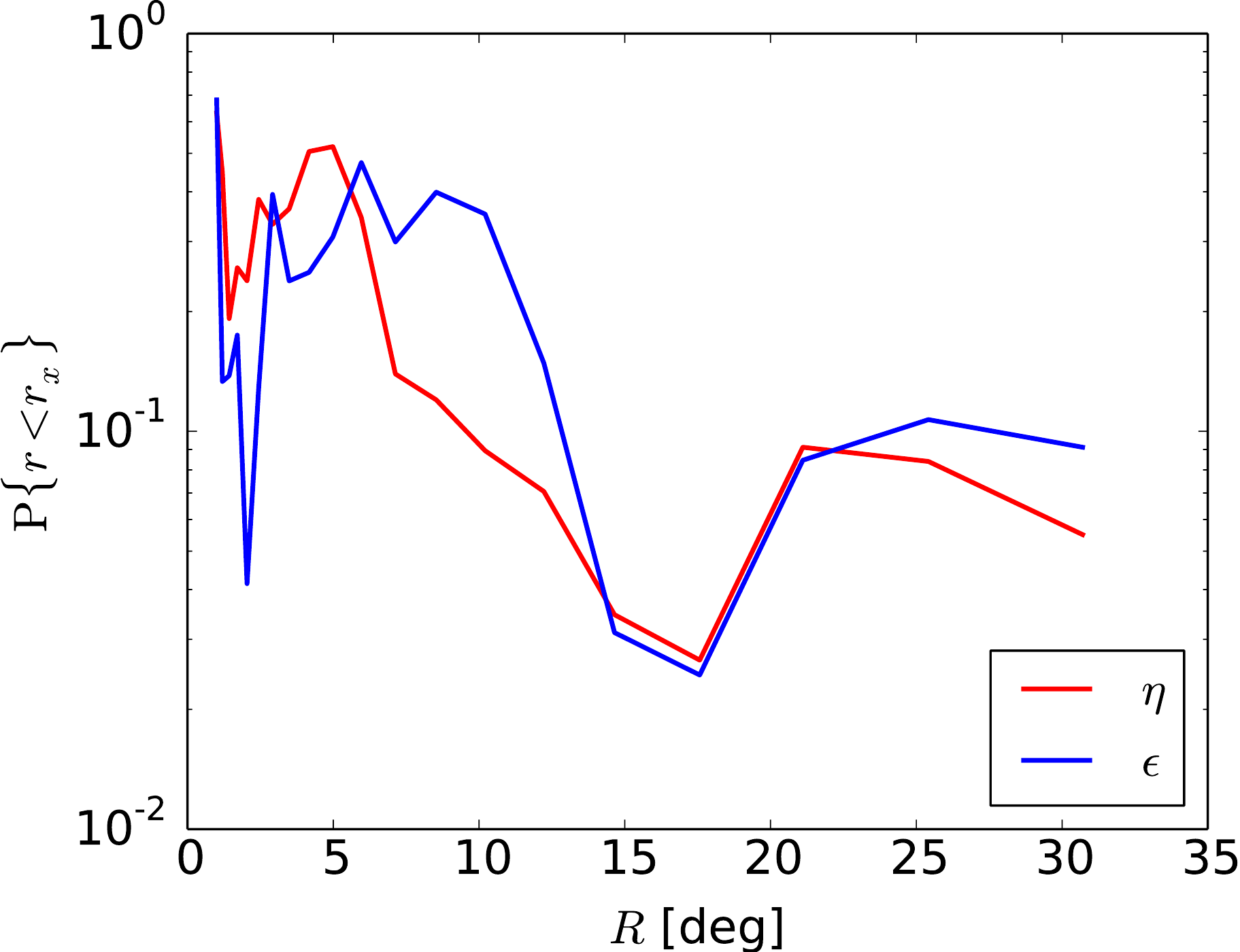}
\includegraphics[scale=0.39]{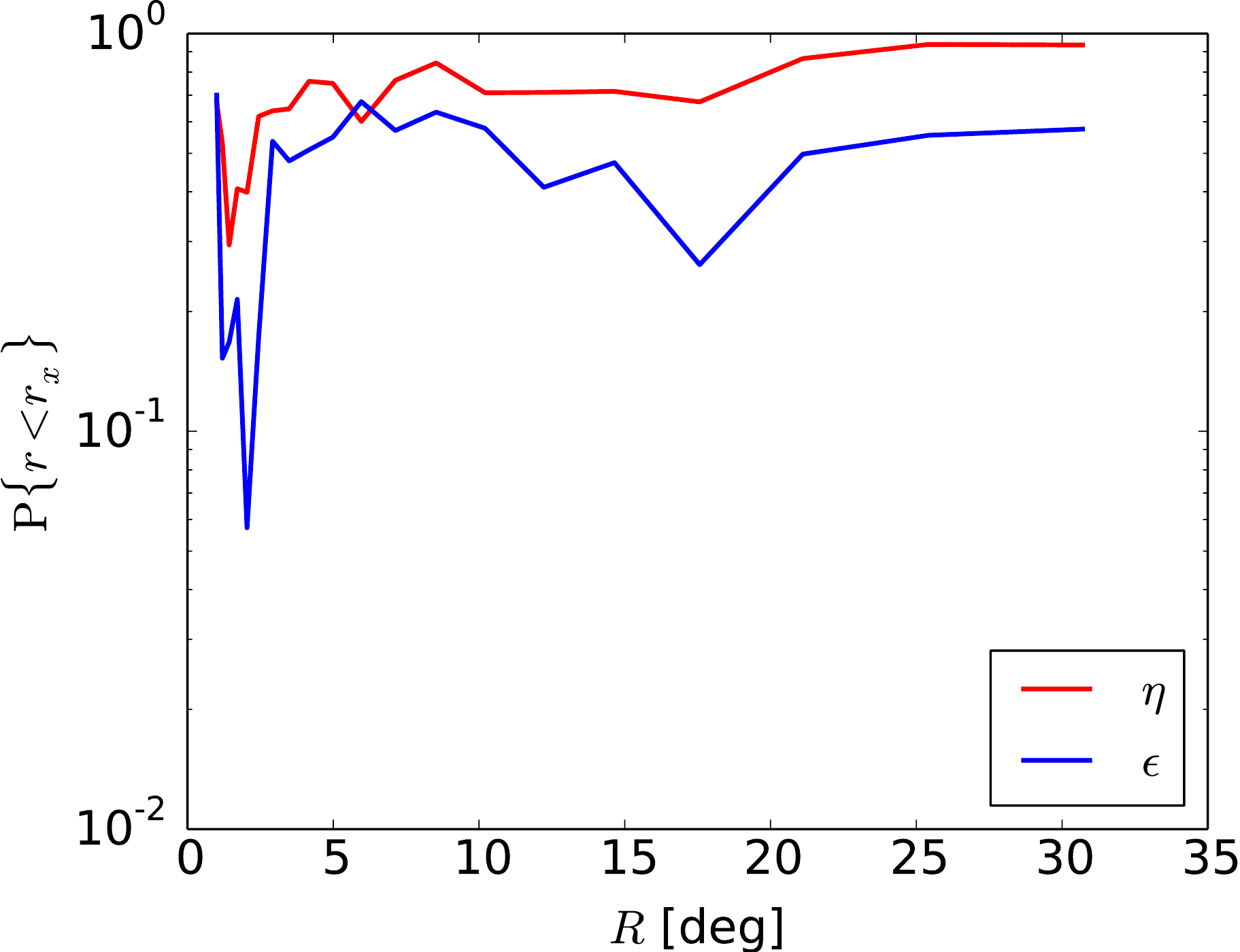}
\end{center}
\caption{Lower tail probability of $r_\eta$ and $r_\epsilon$ as a
  function of the scale, derived from the Planck SEVEM map. Whilst in
  the left figure the corresponding covariances are calculated from
  the theoretical fiducial model, the probabilities in the right
  figure are computed using the estimated covariance from the data.}
\label{fig:pvalue_r_eta_eps}
\end{figure}

In figures~\ref{fig:pvalue_r_nak} and \ref{fig:pvalue_r_eta_eps}, the
upper and lower tail probabilities of $r_x$ for each derivative are
represented. Since a low variance in the derivative fields is
observed, there is a preference for small values of the lower tail
probability, especially at the largest scales. For instance, at
$R=30^\circ$ the probability of having a value of $r_\nu$ lower than
the observed value is about $3\%$ in both SEVEM and SMICA data.  The
low variance is also manifested in $\eta$ and $\epsilon$, in which
case the lower tail probability is below $3\%$ for $R=18^\circ$.
Additionally in figures \ref{fig:pvalue_r_nak} and
\ref{fig:pvalue_r_eta_eps}, in order to take into account this
anomalous variance in the data, the statistics $r_x$ are calculated
using the observed covariance, instead of the covariance obtained from
the theoretical fiducial model. In this case, the anisotropy
introduced by the mask is also modelled using eq.~(\ref{eqn:hatcov}),
but replacing the matrix $\mathbf{C}$ by the estimated covariance from
the data. Once this correction for the low variance is done, the
probabilities of $r_\eta$ and $r_\epsilon$ are within the $2\sigma$
limits, while some values of $r_\nu$ and $r_\kappa$ with upper tail
probabilities $<5\%$ are found.

The statistical deviation caused by the Cold Spot is observed at
$R=5^\circ$ as a decrement in the upper tail probability of
$r_\kappa$. Since we are filtering the temperature with a Gaussian,
the curvature map of the smoothed field is equivalently calculated by
applying the Spherical Mexican Hat Wavelet
(SMHW) \cite{martinez2002}, and therefore, the study of $\kappa$ at
different scales is equivalent to a multiscale analysis using the
SMHW. In previous works \cite{cruz2005} with the SMHW, the Cold Spot
is characterized as the extreme value at $5^\circ$, which causes a
deviation in the skewness at this scale. When the upper tail
probability is computed using the theoretical fiducial model, the CS
represents a relatively likely event with a probability of $6$-$8\%$,
but if the low variance correction is done and the CS fluctuation is
normalized by the observed variance instead of the theoretical one,
this probability falls to $<3\%$, as previously reported in
\cite{cruz2005} and \cite{planck162015}. Consequently, there is a
statistical connection between the CS and the low variance anomaly,
being a rare event having a large fluctuation as the CS in a field
with such a low variance. Besides the curvature field, the combined
analysis of $\nu$ and $\kappa$, where the correlation between them is
taken into account, also presents a deviation at the scale of the CS
and above. In particular, a fluctuation with similar significance to
the CS ($p$-value about $1.2\%$) at the scale of $R \approx 10^\circ$
is observed. At this particular location in the sky, there is no a
extremum in the temperature field filtered with that scale, since the
gradient is different than zero. However, at smaller scales, there is
a hot spot at the same location. This peak and the Cold spot are the
most prominent extrema in the $\kappa$ field at the scale $R \approx
5^\circ$, which were previously identified in the SMHW analysis with a
smaller probability in \cite{planck162015}. It is important to notice
that these deviations may be caused by the low variance of the CMB
temperature field at large scales, since when the analysis is
performed comparing with the fiducial theoretical model, these
fluctuations are less significant (see figure~\ref{fig:pvalue_r_nak}).

\begin{figure}
\begin{center}
\includegraphics[scale=0.34]{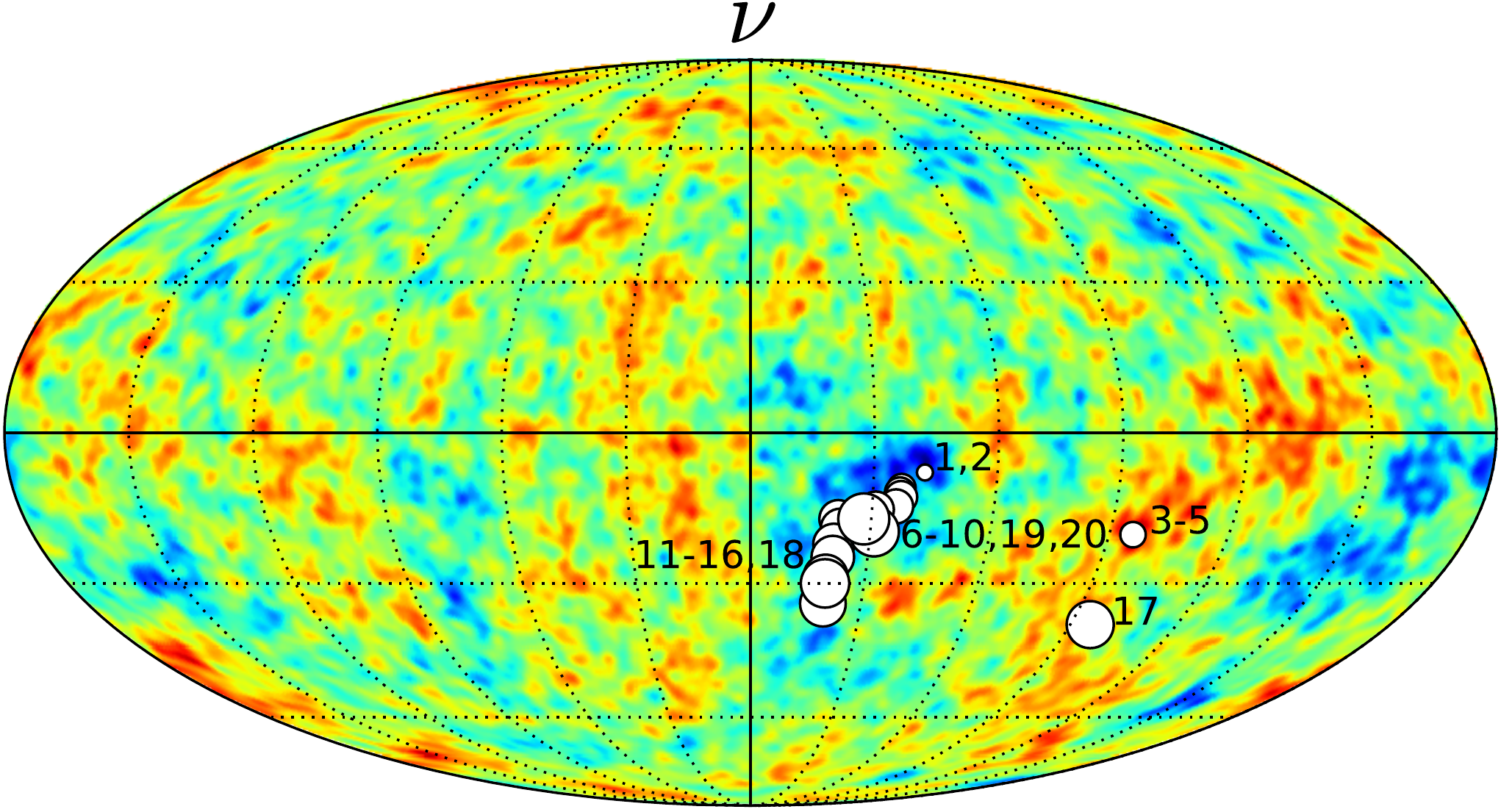}
\includegraphics[scale=0.34]{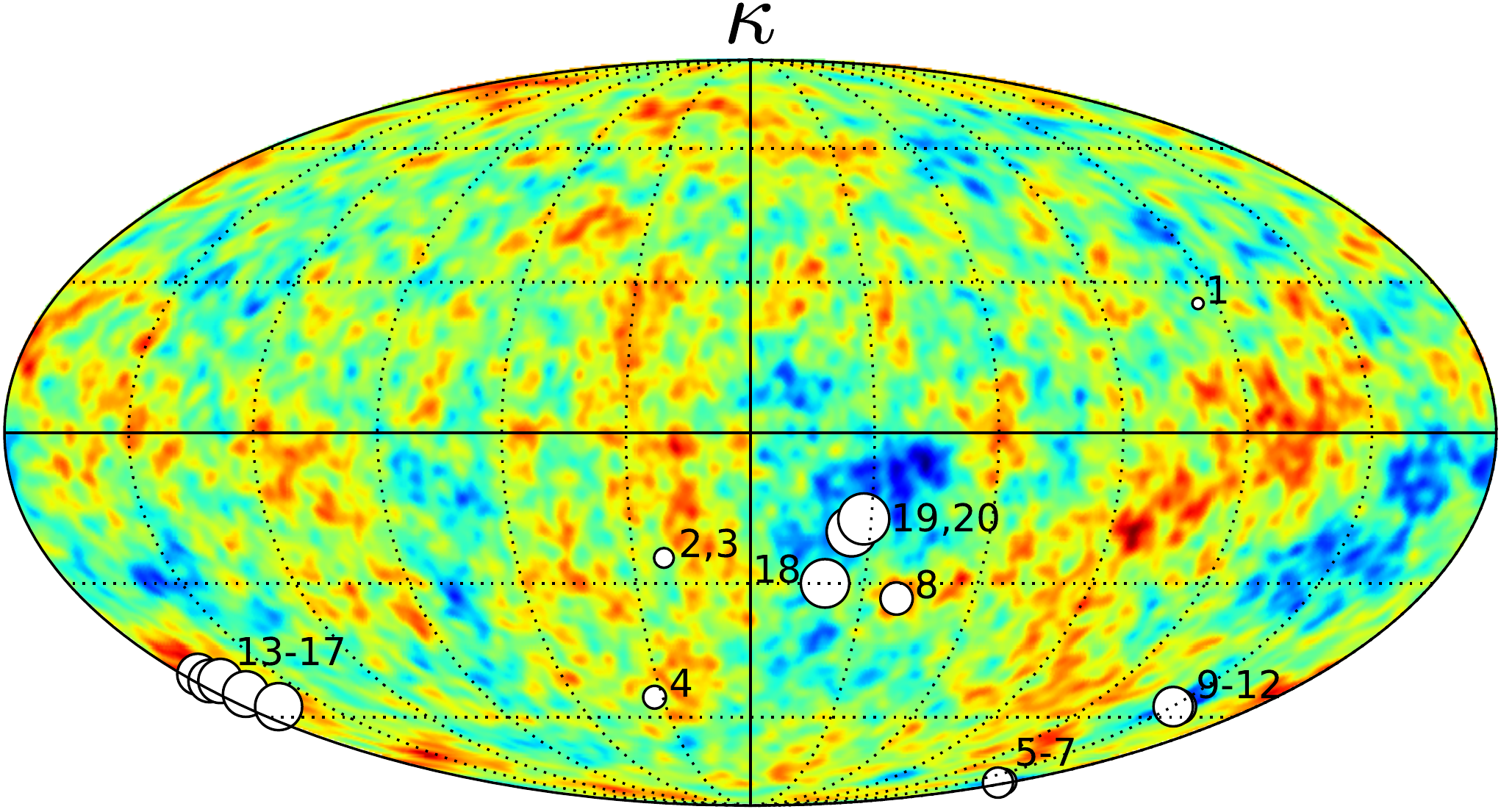}
\includegraphics[scale=0.34]{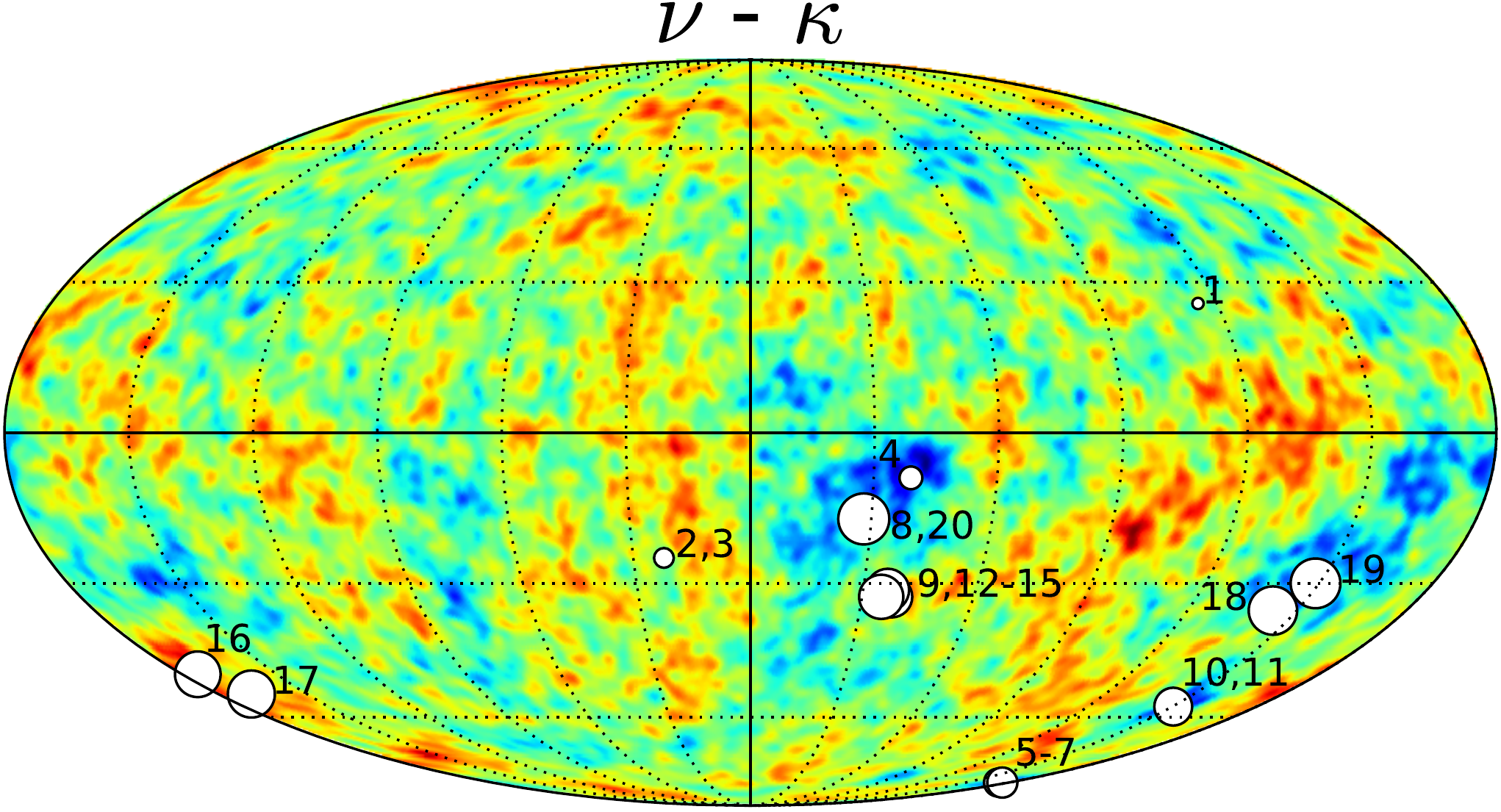}
\includegraphics[scale=0.34]{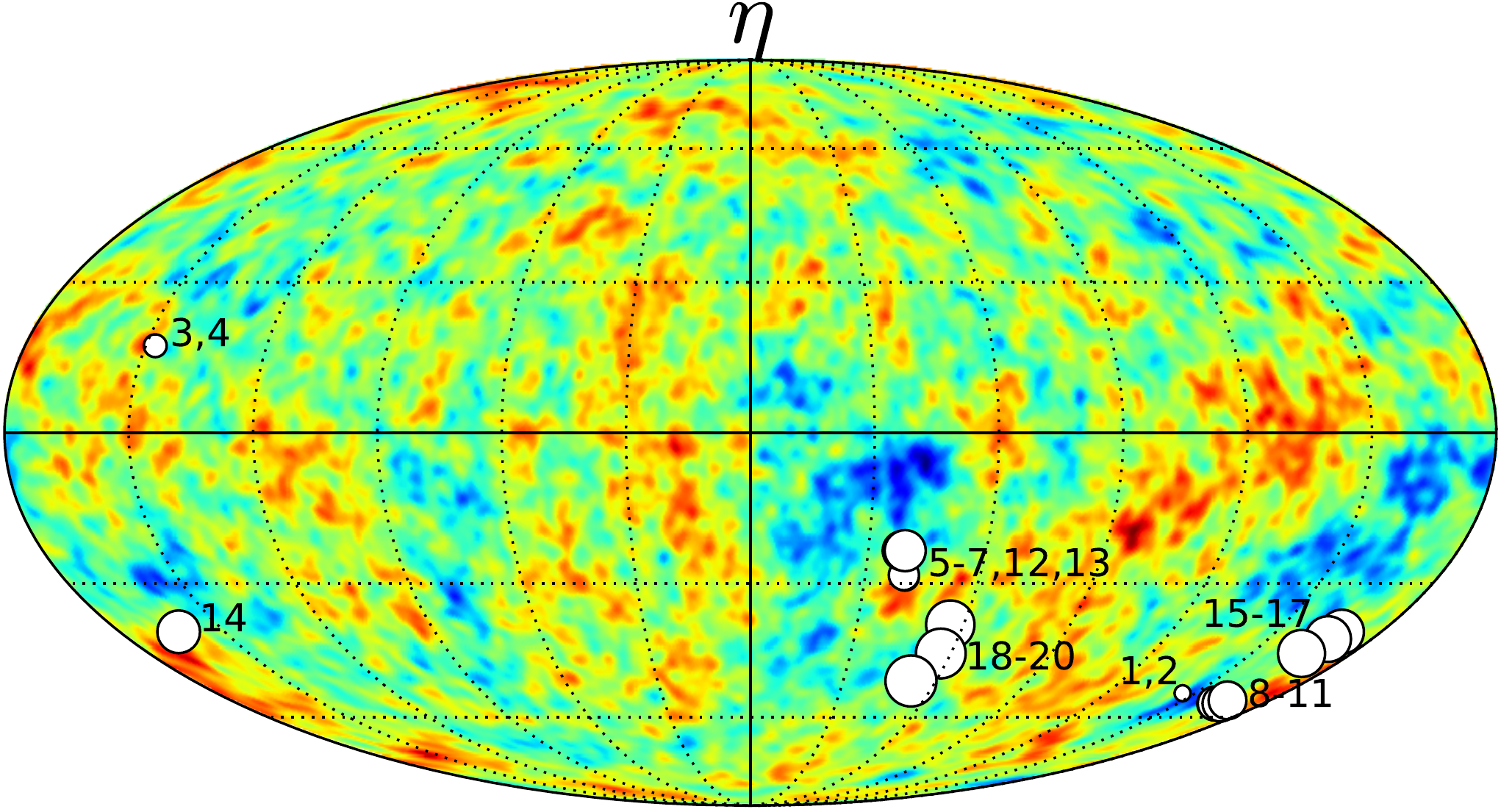}
\includegraphics[scale=0.34]{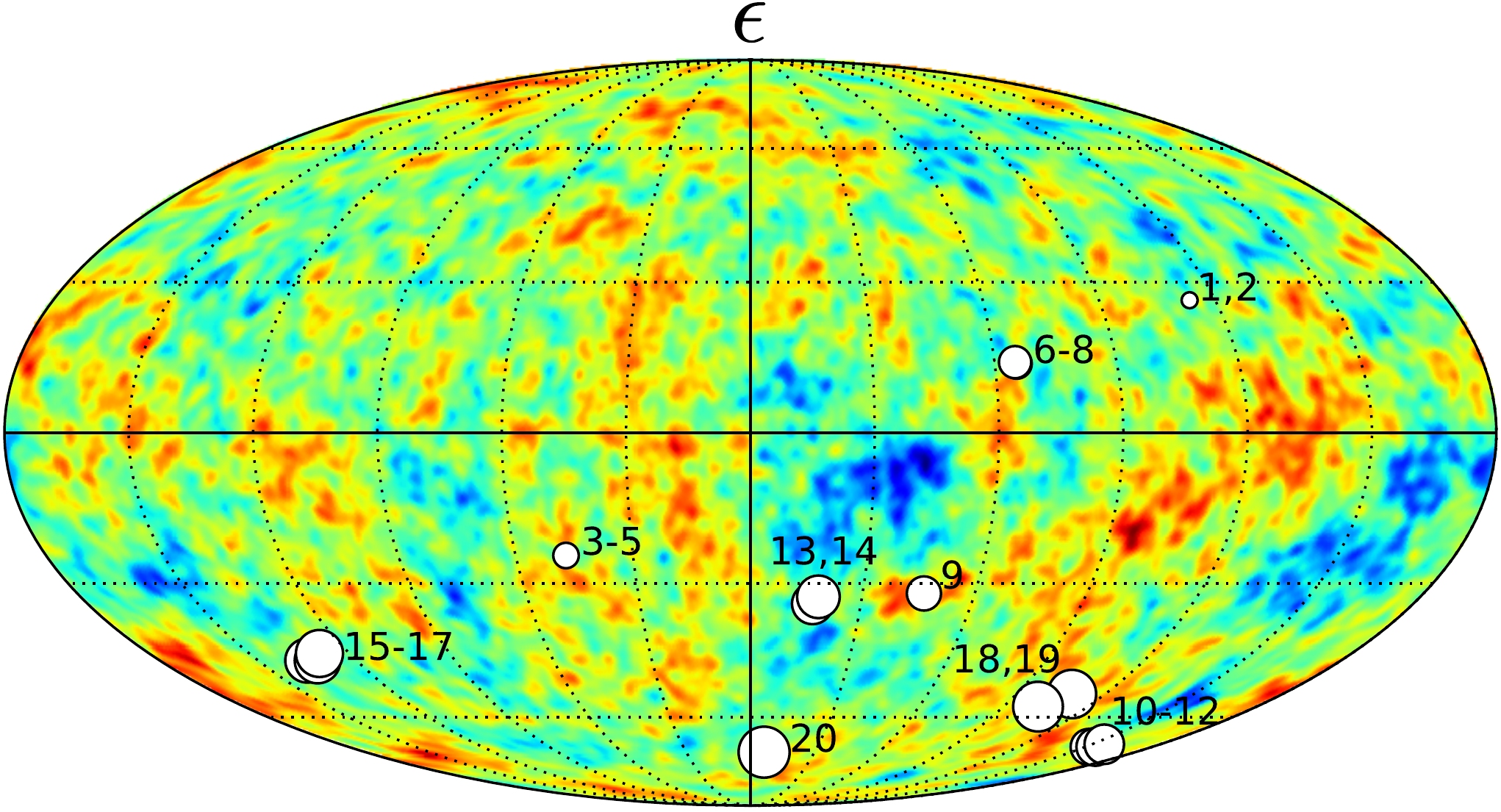}
\end{center}
\caption{Locations of the maximum values of $r_x(p)$ for the
  derivatives fields at different scales. The locations are
  represented by white circles, whose size is proportional to the
  scale $R$ considered. The field corresponding to each map is
  indicated in the title. The CMB temperature field corresponds to the
  Commander Planck map \cite{planck092015} smoothed by Gaussian with
  $R=1^\circ$.}
\label{fig:maxloc}
\end{figure}

The locations of the maximum values of $r_x(p)$ for different scales
are represented in figure~\ref{fig:maxloc}. It is possible to see
that, with the exception of the eccentricity $\epsilon$, all the
maxima lie in the Galactic southern hemisphere, and most of them in
the particular quadrant coinciding with the CMB power asymmetry
\cite{eriksen2004,Hansen2009}. It is important to remark that, in the
case of the temperature $\nu$, all the maxima for scales $R>1^\circ$
are located in two of the largest spots in the sky, one cold and other
hot. On the other hand, the gradient and the second order derivatives
trace other large scale features (e.g. the Cold Spot, traced by the
$\kappa$ field). In the case of $\eta$ and $\kappa$, the maximum
deviation from the standard model are located near to the largest
peaks observed in the temperature map, whereas deviations of the
eccentricity tensor $\epsilon$ are spread along the field without any
particular clustering around the largest structures. The excess of
clustering of the $\nu$ maxima compared with high-order derivatives is
caused by the particular scale dependence of the derivatives, which
introduces extra $\ell$ factors, obtaining, in this way, less
correlated extrema.

\section{Directional analysis}
\label{sec:dir_analysis}

\begin{figure}
\begin{center}
  \includegraphics[scale=0.34]{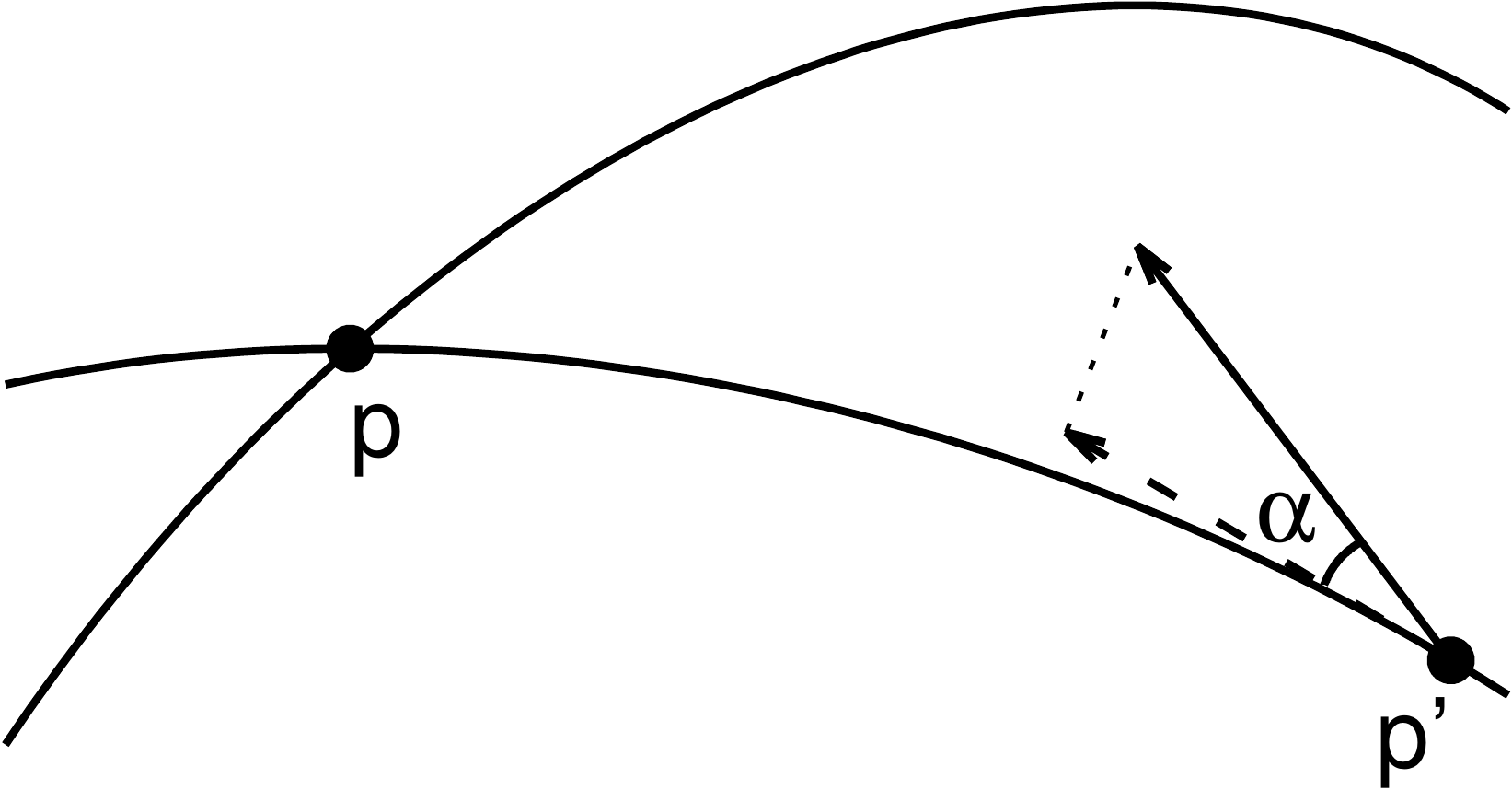}
  \hspace{1cm}
  \includegraphics[scale=0.34]{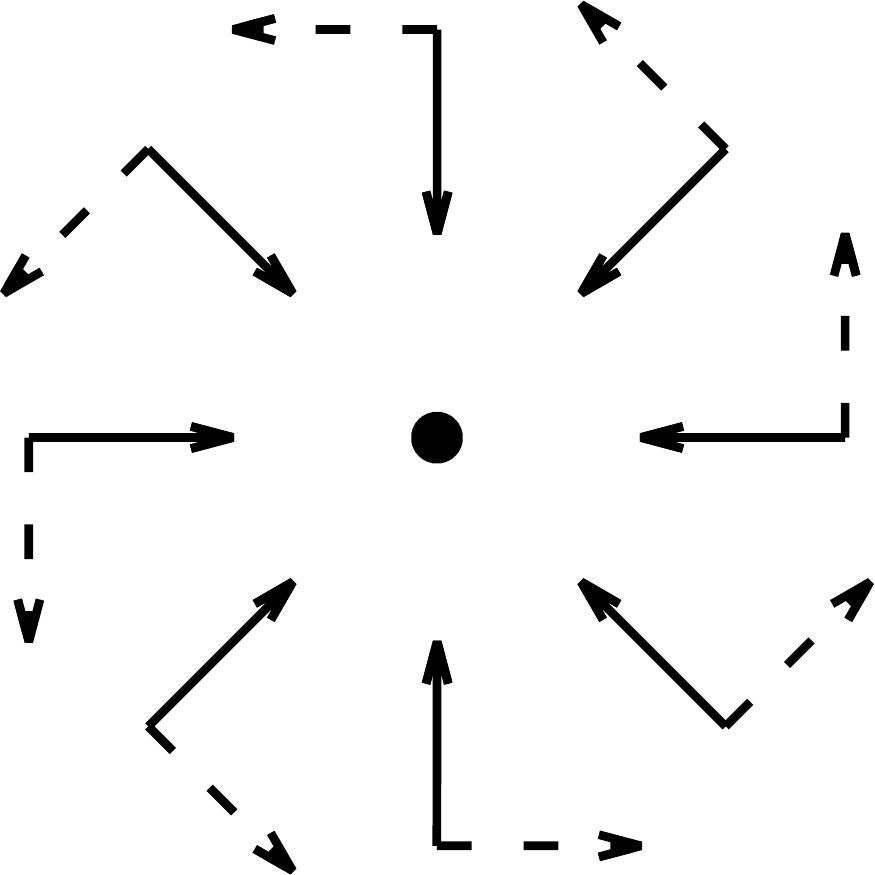}
  \hspace{0.5cm}
  \includegraphics[scale=0.34]{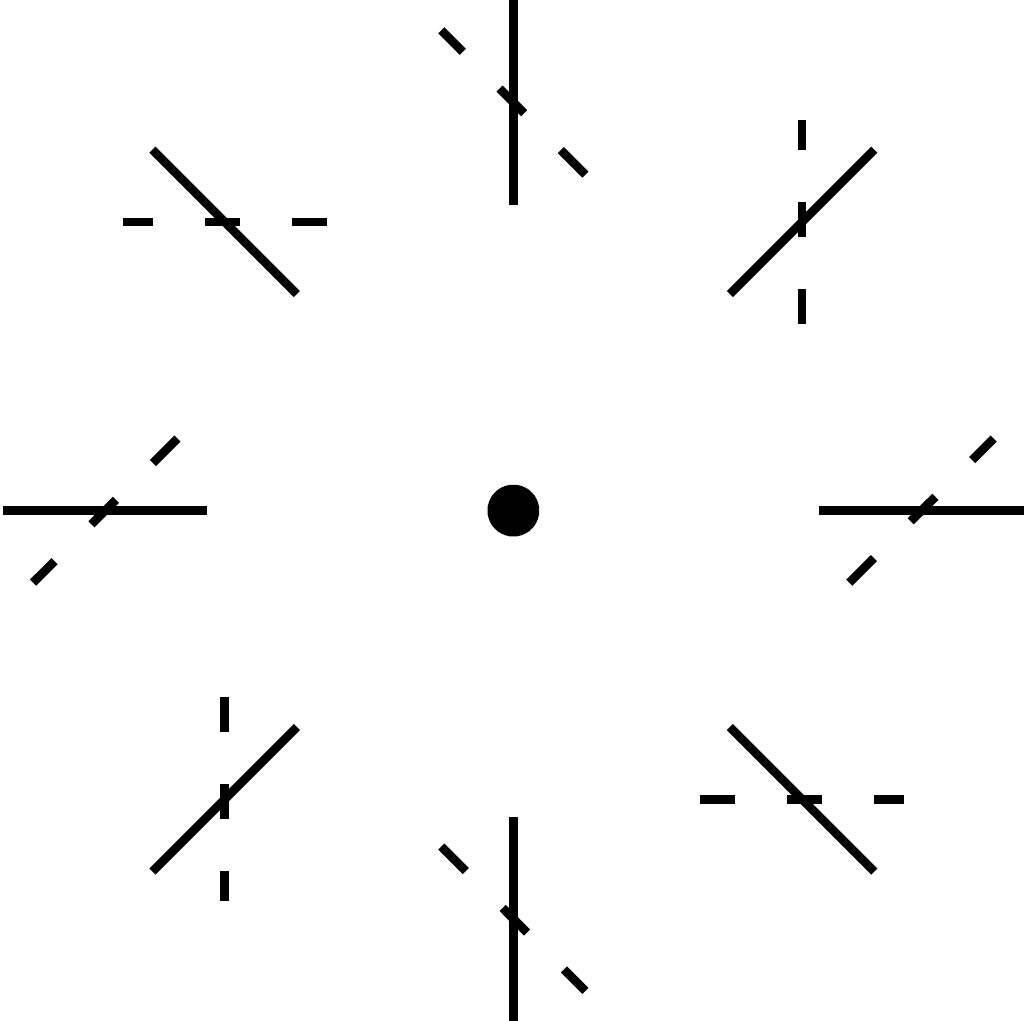}
\end{center}
\caption{\emph{Left:} projection of a vector field ($s=1$) along the
  geodesic connecting $p$ and $p^\prime$ as it is considered for the
  definition of the directional asymmetry estimator in
  eq.~(\ref{eqn:dir_est}). \emph{Right:} the two pictures represent
  the contribution of the spinor ${}_s\zeta$ field to the real and
  imaginary parts of ${}_s\bar{\zeta}(\mathbf{x})$. For simplicity,
  only the particular cases of $s=1$ (vectors) and $s=2$ (headless
  vectors) are depicted in the figure. The reference point
  $\mathbf{x}$ used for the projection is indicated at the centre of a
  small flat patch of the sphere. The spinor components which have
  even parity (solid lines) contribute to the real part of
  ${}_s\bar{\zeta}$, whereas its imaginary part measures the
  contribution of the odd parity components (dashed lines). A change
  in the sign of ${}_s\bar{\zeta}$ can be seen a rotation of $\pi/s$
  in the spinor field.}
\label{fig:dir_proj}
\end{figure}

The gradient and the eccentricity tensor are spinorial quantities with
non-zero spin, and therefore they have directional dependence which
can be used to study alignment directions and the local isotropy of
the field. In this section, an estimator of preferred directions in
the sky for spinorial quantities is introduced. Given a particular
point $p$ on the sphere, we can construct all the geodesics connecting
it with any other arbitrary point (see
figure~\ref{fig:dir_proj}). This system of geodesics define a
particular directional scheme associated to $p$, which can be used to
project the spinorial field along this geodesics. The projected field
is averaged using a particular weight function $W$, which depends on
the size of the region around the point $p$ where the isotropy is
tested. If this process is repeated for all the possible points on the
sphere, we get an estimator depending on the point $p$ which define
the anisotropy direction. For instance, in the case of the gradient,
this estimator measures a possible excess of vectors pointing to a
particular location on the sphere. This concept can be generalized for
arbitrary spin quantities, maintaining a similar interpretation. In
the case of the eccentricity tensor, which is a $2$-spinor, this
estimator indicates the existence of predominant directions where the
local elongations of peaks are oriented. A similar analysis of the
alignment of the eccentricity tensor (equivalent to the steerable
wavelet basis formed by the second Gaussian derivatives) based on the
intersection of great circles was applied to WMAP data in
\cite{wiaux2006,vielva2006}.

Concretely, given a spinorial field ${}_s\zeta$ with spin $s$ defined
in the Galactic coordinates, the isotropy estimator associated to the
direction defined by the point $\mathbf{x}$ is:
\begin{equation}
{}_s\bar{\zeta}(\mathbf{x}) = \int \mathrm{d}^2\mathbf{y}
\ W(\mathbf{x}\cdot\mathbf{y}) {}_s\zeta(\mathbf{y})
e^{is\alpha(\mathbf{x},\mathbf{y})} \ ,
\label{eqn:dir_est}
\end{equation}
where $\alpha(\mathbf{x},\mathbf{y})$ is the angle between the
Galactic north direction and the geodesic connecting the points
$\mathbf{x}$ and $\mathbf{y}$. The function $W$ in this integral
weights the contribution of the spinor at $\mathbf{y}$ in the
anisotropy direction defined by $\mathbf{x}$. In this way, the
function $W$ has information about the region around the point
$\mathbf{x}$ where the directional analysis performed. It is possible
to see that the directional estimator ${}_s\bar{\zeta}(\mathbf{x})$ is
a scalar field, which can be expanded in terms of the spin zero
spherical harmonics. In general, this field is complex, and its real
and imaginary parts correspond to the projection of different parity
components of the spinor (see figure~\ref{fig:dir_proj}).

The estimator in eq.~(\ref{eqn:dir_est}) can be written in the
spherical harmonic space as:
\begin{equation}
{}_s\bar{\zeta} (\mathbf{x}) = \sum_{l=|s|}^\infty \sum_{m=-\ell}^\ell D_\ell^s
\sqrt{\frac{(\ell-|s|)!}{(\ell+|s|)!}} \ {}_s\zeta_{\ell m} \ Y_{\ell m}
(\mathbf{x}) \ ,
\label{eqn:zeta_bar}
\end{equation}
where ${}_s\zeta_{\ell m}$ are the spherical harmonics coefficient
associated to the spinor ${}_s\zeta(\mathbf{x})$. The spin dependent
quantities $D_\ell^s$ in eq.~(\ref{eqn:zeta_bar}) are given by:
\begin{equation}
D^s_{\ell} = \sum_{\ell^\prime=0}^\infty M_{\ell\ell^\prime}^s W_{\ell^\prime}
\ ,
\end{equation}
where $W_\ell$ represents the coefficients of the filter $W$ in
eq.~(\ref{eqn:dir_est}) in harmonic space and the coupling matrix
$M_{\ell\ell^\prime}^s$ is given by
\begin{equation}
M_{\ell\ell^\prime}^s = s (-1)^s \left( 2 \ell^\prime +1 \right)
\sqrt{\frac{(\ell+|s|)!}{(\ell-|s|)!}}
\sum_{L=|\ell-\ell^\prime|}^{\ell+\ell^\prime} \left( 2L+1 \right)
\sqrt{\frac{(L-|s|)!}{(L+|s|)!}} \left( \begin{array}{ccc} \ell &
  \ell^\prime & L \\ 0 & 0 & 0 \\ \end{array} \right)
\left( \begin{array}{ccc} \ell & \ell^\prime & L \\ s & 0 & -s
  \\ \end{array} \right) A_L^s \ .
\end{equation}
In this expression, the numbers $A_\ell$ are related with integral of
the spin-weighted spherical harmonics, which are defined in the
Appendix~\ref{app:sh_integral}.

In the particular case of the gradient and the eccentricity tensor, it
is possible to see that the estimator in eq.~(\ref{eqn:dir_est}) can
be obtained by filtering the temperature map with the filters
$D_\ell^s$:
\begin{subequations}
\begin{equation}
\bar{\eta}(\mathbf{x}) = \sum_{\ell=1}^\infty \sum_{m=-\ell}^\ell
D_\ell^1 a_{\ell m} Y_{\ell m} (\mathbf{x}) \ ,
\end{equation}
\begin{equation}
\bar{\epsilon}(\mathbf{x}) = \sum_{\ell=2}^\infty \sum_{m=-\ell}^\ell
D_\ell^2 a_{\ell m} Y_{\ell m} (\mathbf{x}) \ .
\end{equation}
\end{subequations}
Notice that the directional estimators for the derivatives
$\bar{\eta}(\mathbf{x})$ and $\bar{\epsilon}(\mathbf{x})$ are real
scalar fields. Moreover, since they have a linear dependence on the
temperature field, it is assumed that they are Gaussianly distributed,
which simplify the calculations and the subsequent statistical
analysis.

\begin{figure}
\begin{center}
  \includegraphics[scale=0.39]{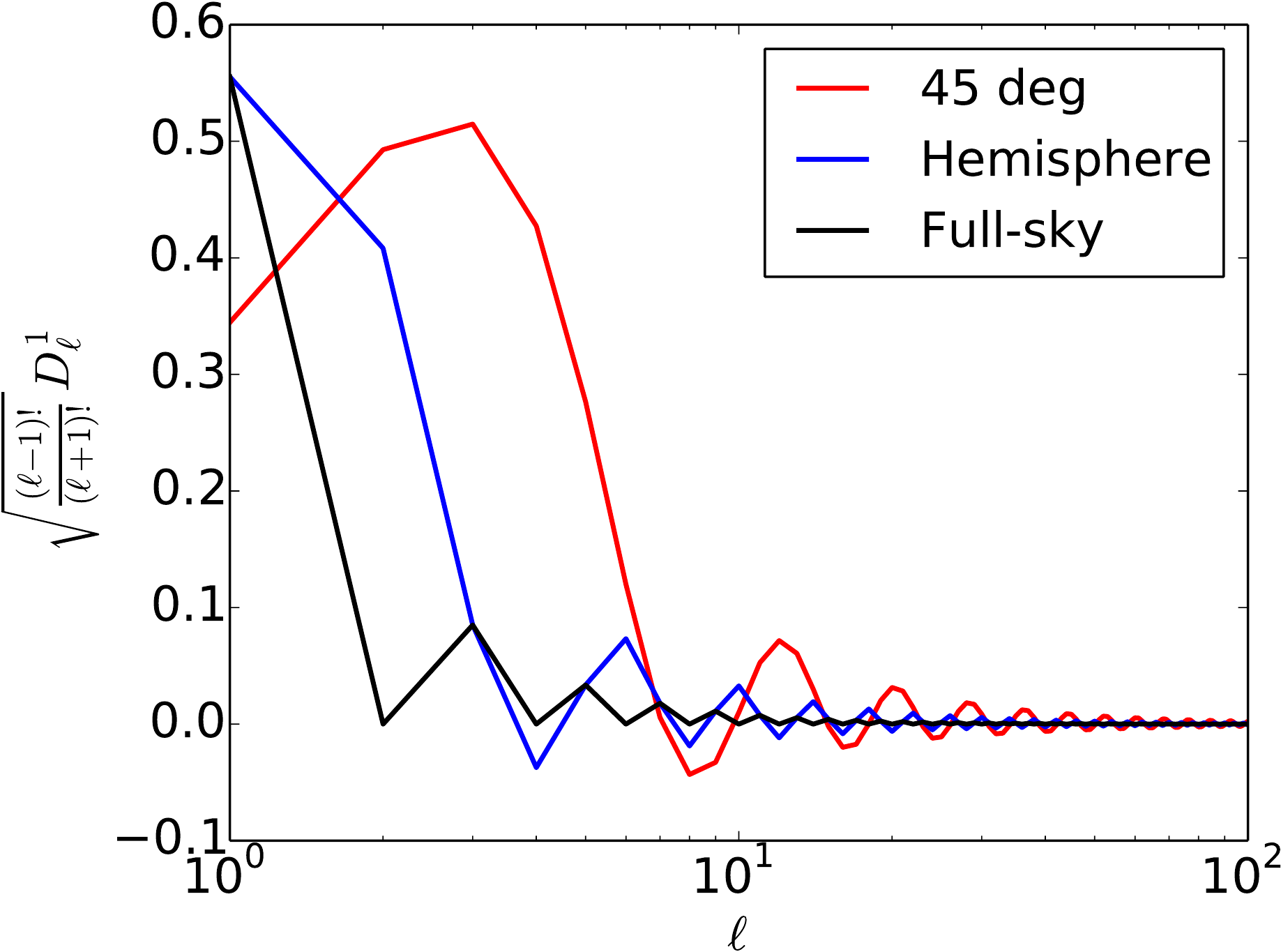}
  \includegraphics[scale=0.39]{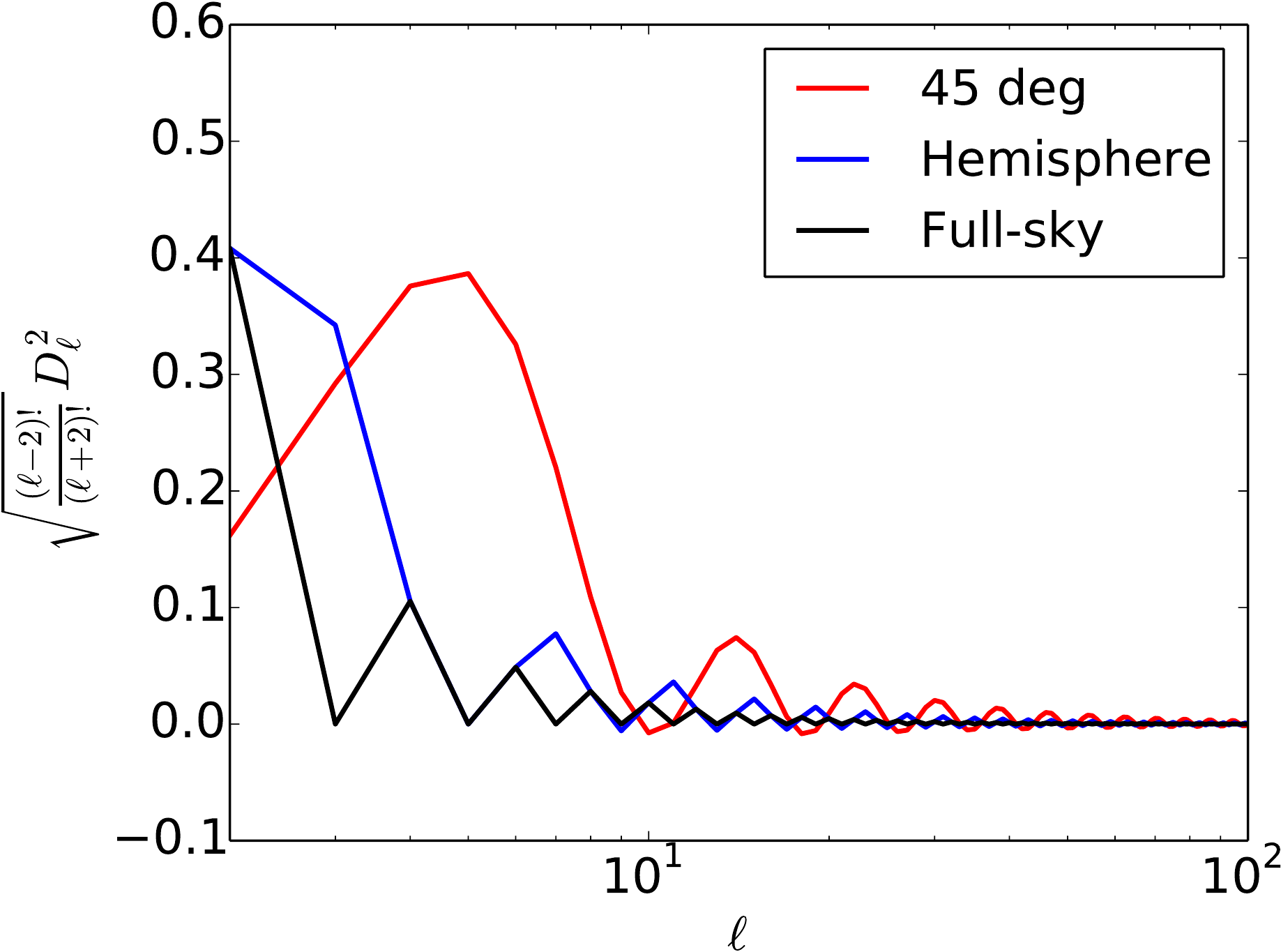}
\end{center}
\caption{Filters $D_\ell^s$ for $s=1$ (left) and $s=2$ (right) used in
  the analysis of the directionality of spinors. Different curves
  correspond to different averaged area in the estimator.}
\label{fig:dir_filters}
\end{figure}

In this work, the weight function $W$ used for averaging of the
projected spinor in eq.~(\ref{eqn:dir_est}) is assumed to be a disc
centred at the point $\mathbf{x}$\footnote{In the harmonic space, the
  weight function $W$ corresponding to a disc with radius $\theta$ is
  given by \[ W_\ell(\mu) = \begin{cases} - \sqrt{\frac{1+\mu}{1-\mu}}
    \frac{P_\ell^1(\mu)}{\ell(\ell+1)} \ , & \ell \neq 0 \\ 1 \ , &
    \ell = 0 \end{cases} \ , \] where $\mu = \cos \theta$ and
  $P_\ell^1(\mu)$ is the associated Legendre polynomial with
  $m=1$.}. Particularly, discs with different radius are considered:
$180^\circ$ (full-sky average), $90^\circ$ (one hemisphere) and
$45^\circ$. The resulting filters $D_\ell^s$ for these three cases are
represented in figure~\ref{fig:dir_filters}, for $s=1$ and $s=2$. The
asymmetry in the multipoles with different parity, which is more
evident in the filters with the largest averaging window, is a
consequence of the transformation rules of the spinors under the
rotation group. For instance, when a $s$-spinor is projected and
averaged over the full sphere, the multipoles with parity different
from $s$ vanish because the field ${}_s\bar{\zeta}(\mathbf{x})$ has
well-defined parity given by $(-1)^s$. More precisely, in this
particular case, the value of ${}_s\bar{\zeta}$ at the point
$\mathbf{x}$ and at its antipode $-\mathbf{x}$ are related by a
rotation of $\pi$ radians, which introduces a factor $-1$ depending on
the parity of the spinor (odd for the vector $\eta$ and even for
$2$-spinors $\epsilon$).

\begin{figure}
\begin{center}
\begin{tabular}{cc}
\multirow{2}{*}[10em]{\includegraphics[scale=0.60]{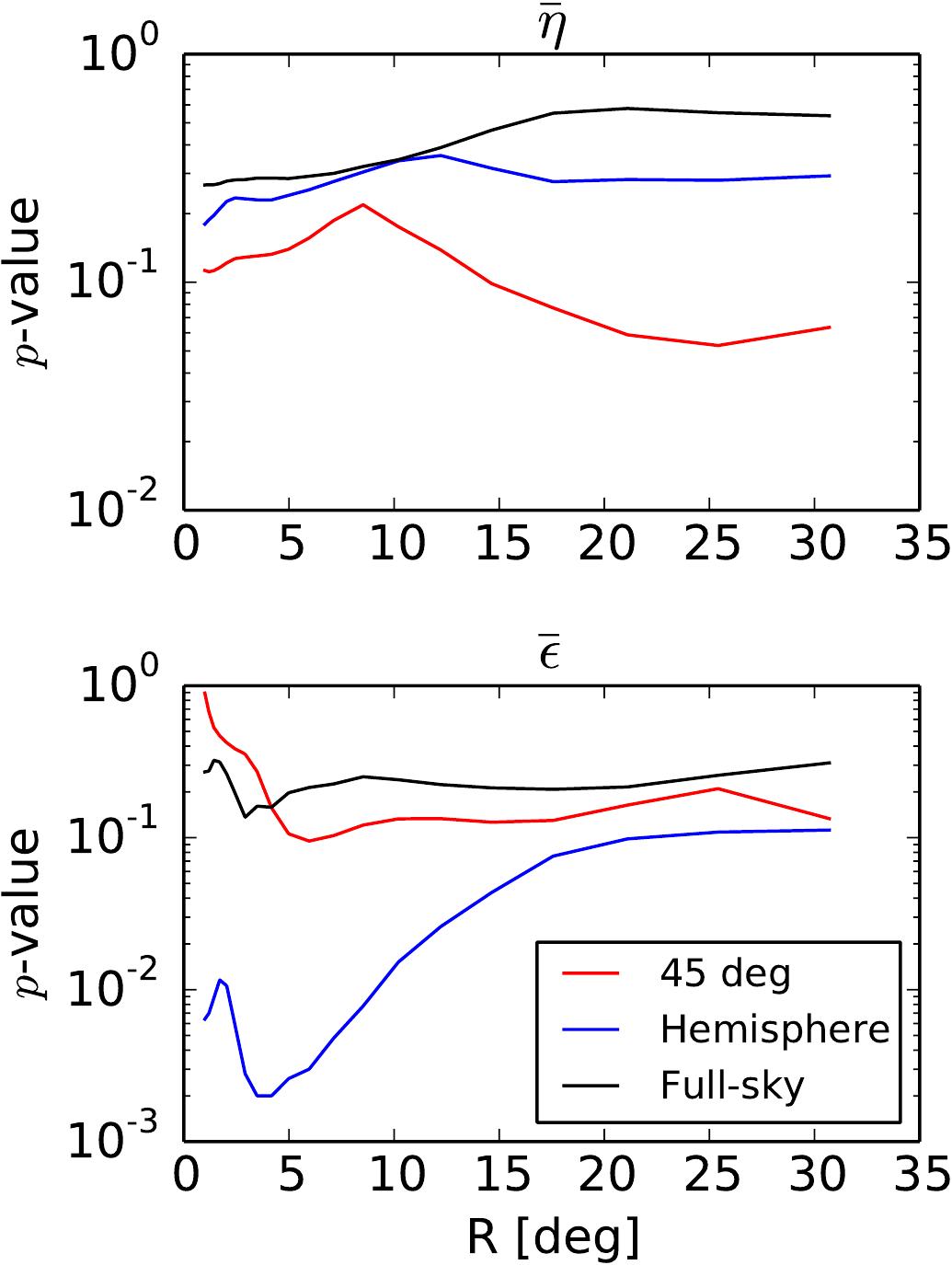}}
& \includegraphics[scale=0.35]{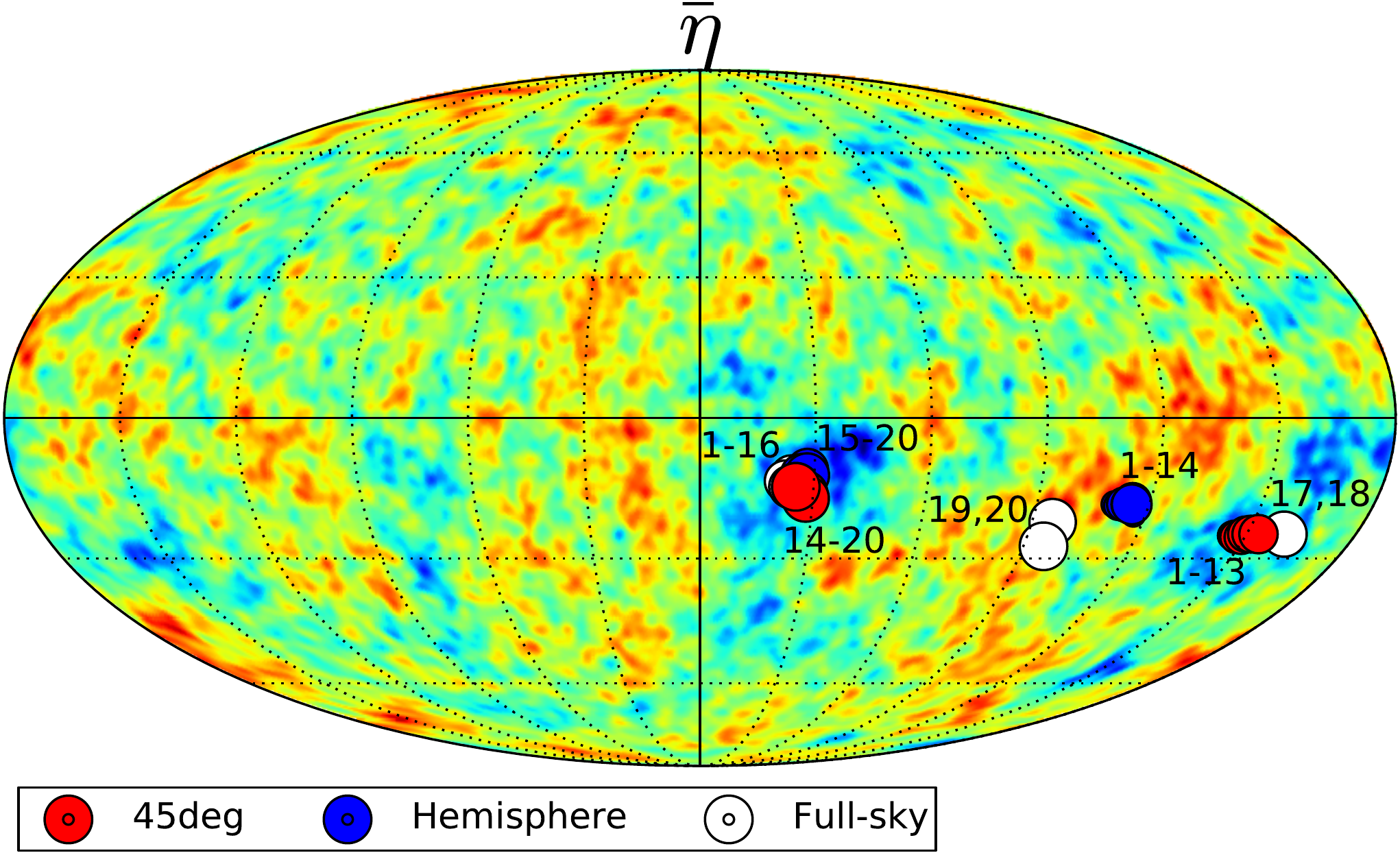} \\
& \includegraphics[scale=0.35]{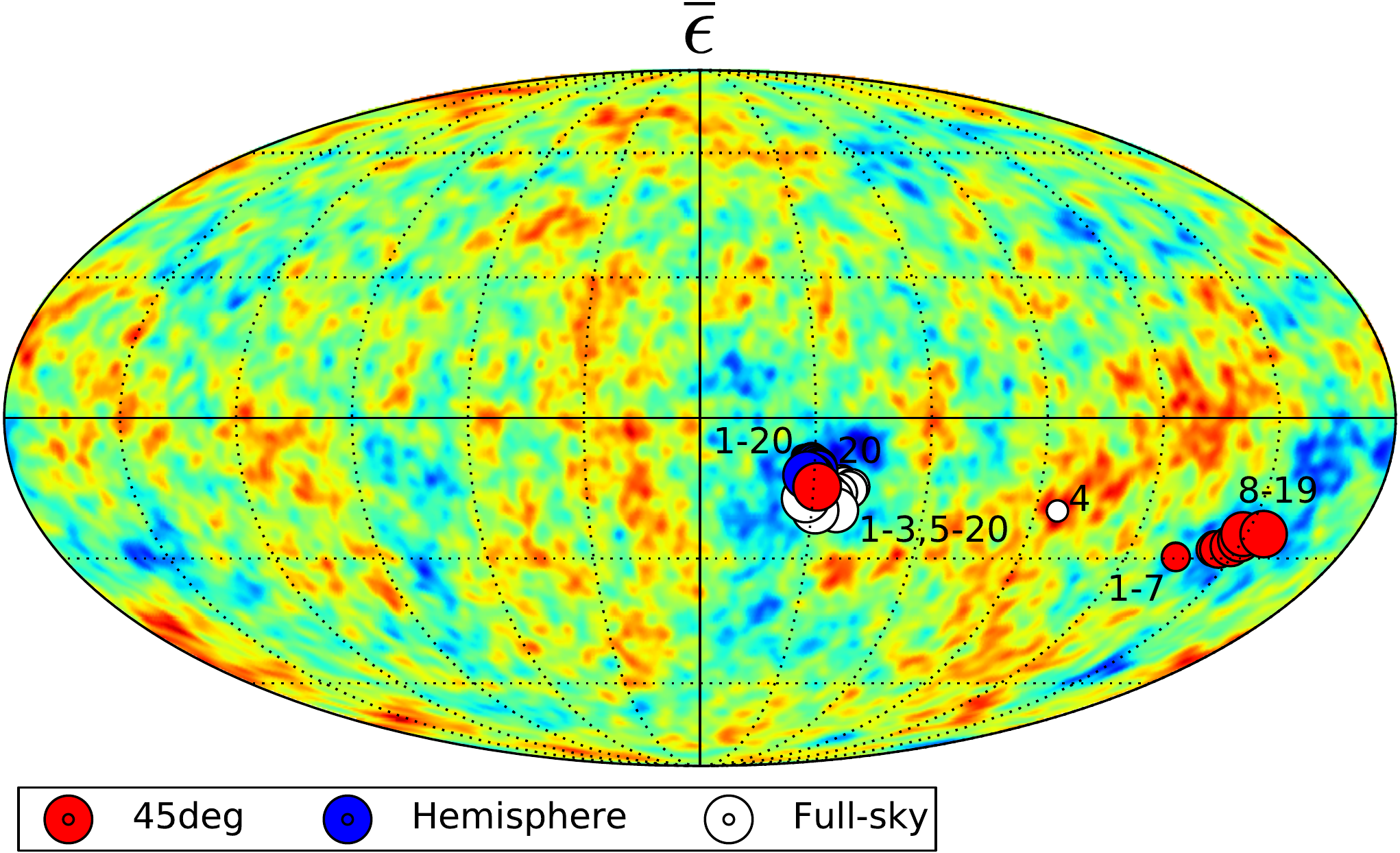} \\
\end{tabular}
\end{center}
\caption{Directional analysis of the spinors $\eta$ and $\epsilon$
  using the three cases considered: full-sky, hemispherical and
  $45^\circ$ averages. \emph{Left:} the probabilities of finding a
  value of $\bar{\eta}$ (upper figure) and $\bar{\epsilon}$ (lower
  figure) as extreme as the maximum observed in the data for different
  scales. \emph{Right:} the locations of these maxima on the CMB map
  are indicated with circles whose size is proportional to the scale
  $R$. The results presented in this figure are obtained from the
  Planck SEVEM map.}
\label{fig:dir_analysis}
\end{figure}

Within this formalism, the points on the sphere which present higher
directional asymmetry in the derivatives correspond to extrema in the
scalar fields $\bar{\eta}(\mathbf{x})$ and
$\bar{\epsilon}(\mathbf{x})$, which, in particular, are a maximum or a
minimum depending on whether the spinor tends to be aligned or
anti-aligned with the direction given by $\mathbf{x}$. From this point
of the study, these extrema are calculated and characterized in the
same way as the analysis performed in Section~\ref{sec:extreme_der},
and with the same considerations for the mask derived in
Section~\ref{sec:pixel_cov}. The results are shown in
figure~\ref{fig:dir_analysis}, where we represent the $p$-value of the
extrema in $\bar{\eta}$ and $\bar{\epsilon}$ observed in the data as a
function of the scale $R$, as well as their corresponding location on
the CMB field. The same three cases represented in
figure~\ref{fig:dir_filters} for the weight function are considered in
the analysis. Whilst the gradient is compatible with the standard
model prediction, the eccentricity tensor has a preferred
directionality when the spinor is averaged over an hemisphere. The
probability of this deviation is only $0.2\%$ for $R=4^\circ$ in the
SEVEM map, whereas this values increases up to $0.9\%$ for the SMICA
data. The corresponding anisotropic direction is located near the
Galactic plane on one of the largest peaks of the CMB (see
figure~\ref{fig:dir_analysis}). However, notice that the main
contribution to this deviation come from a strip between $45^\circ$
and $90^\circ$ from the centre of this structure, since the estimator
$\bar{\epsilon}$ for the $45^\circ$ averaging is within the standard
model limits.

\section{Conclusions}
\label{sec:conclusions}

In this work, the CMB temperature field is analysed by calculating its
derivatives up to second order at different scales. One of the
problems is that the incomplete sky (due to the masking of the
Galactic emission and the point sources) causes a wrong determination
of the derivatives at the border of the mask. In addition, other
systematics appear when a convolution is applied to the masked data,
since the filtering introduces a smearing of the mask border which is
proportional to the filter scale. Therefore, in a multiscale analysis
of the derivative fields, the handling of the mask is important in
order to have a correct characterization of the derivatives. Due to
the fact that the mask breaks the statistical isotropy of the field,
the covariance of the fields depends on the pixel location, following
the geometry given by the particular mask considered. The calculation
of the pixel covariance is achieved in Section~\ref{sec:pixel_cov} by
doing Monte Carlo simulations in an efficient way in order the reduce
the simulation errors. For this purpose, the covariance at a given
pixel is expressed as a linear transformation of the theoretical
pixel-independent covariance, using a Cholesky-like
decomposition. Following this procedure, we have that the temperature
$\nu$ and the local curvature $\kappa$ at a given pixel are correlated
in a way determined by the theoretical fiducial model and the mask
geometry. Besides, the components of the spinorial derivatives (the
gradient and the eccentricity tensor), which are independent in an
isotropic field, are correlated as a consequence of the incomplete
sky.

Once the covariances between the different derivatives components have
been determined at each pixel, an estimator of the theoretical
full-sky covariances is proposed in Section~\ref{sec:der_cov}, which
generalizes the standard maximum likelihood estimator for full-sky
data. A multiscale analysis is performed by calculating these
covariances at different scales, finding that there is a systematic
low variance preference at large scales in all the
derivatives. Regarding the off-diagonal terms, an unusual low
correlation between $\nu$ and $\kappa$ is found when it is compared
with its theoretical prediction. But, on the other hand, this effect
disappears when the correlation term is normalized by the respective
measured variances, indicating that the low correlation is directly
related to the anomalous low variance.

Moreover, the isotropy of the field can be tested by looking at the
variance of the gradient and the eccentricity tensor. If there is no
preferred directions on the sky, the variances of each spinorial
component must be the same, and the correlation between them should
vanish. By comparing these assumptions as a function of the scale, no
deviation from the isotropy is found in the CMB temperature. The
statistical properties of the spinor components depend on the
particular local system of reference used in their description, and
therefore, this result is associated to the $z$ axis of the standard
Galactic coordinates. A more general analysis varying the azimuthal
direction has to be performed in order to conclude that the
derivatives are statistically isotropic.

The possible departure of the data from the standard model is
quantified by looking at the deviation of the extreme values of the
derivatives fields. The procedure consists in comparing the measured
value of the derivative with the pixel covariance calculated in
Section~\ref{sec:extreme_der} using a $\chi^2$ test. The deviations of
the extreme values are quantified as the tail probability of finding
that value in one realization. In this analysis, the observed low
variance in the data has an important role in the determination of
this quantity, which causes that the extrema have particularly small
values. In order to correct by this effect, pixel covariances which
takes into account the mask geometry as well as the observed low
variances of the derivative fields are introduced. Repeating the
analysis with these modified covariances, the anomaly in the values of
the extrema disappears in all the derivative fields, with the
exception of $\kappa$, where deviations associated to the Cold Spot
\cite{cruz2005} and other large scale fluctuations are observed.  In
addition, a deviation at the scale $R = 10^\circ$ is highlighted in
the combined analysis of the $\nu$ and $\kappa$ whose $p$-value is
comparable with the Cold Spot. The spatial location of the extrema is
concentrated in the southern ecliptic hemisphere, a region which
appear to be anomalous in other estimators of the statistical
isotropy, as the dipole modulation \cite{hoftuft2009}. It is possible
to conclude from these results that the significance of these
anomalies in the CMB temperature at large scales may be related of the
low variance of the field. When these deviations are referred to the
variance calculated from the theoretical model instead of the ones
obtained from the data, the compatibility of the deviations increase
to a probability of $6\%$.

Finally, in Section~\ref{sec:dir_analysis}, an estimator of the local
isotropy of the field based on the geodesic projection is developed
for spinorial quantities. The mathematical formalism can be reduced to
the application of a given kernel to the spherical harmonic
coefficients, which is a function of the particular spin of the
quantity considered. This directional analysis depends on the sky area
used for averaging the projected spinor, which allows an analysis of
isotropy at different scales. Since we are interested in the large
scales, we consider three cases in our study: full-sky, one hemisphere
and $45^\circ$ averaged areas. As in the previous section, deviations
from the standard model are characterized by the directions of maximum
anisotropy. The results indicate that these directions correspond to
the largest structures observed in the CMB temperature. In particular,
it is observed a deviation whose $p$-value is $\approx 0.2$-$0.9\%$
which is centred in one of the largest peaks near the Galactic
plane. In a future work, this analysis of the isotropy of the CMB
based on spinorial quantities will be generalized to the polarization
spinor.

\appendix

\section{Integrals of the spin-weighted spherical harmonics}
\label{app:sh_integral}

The spatial average of the spin-weighted spherical harmonics are given
by:
\begin{equation}
\frac{1}{4\pi} \int \mathrm{d}^2\mathbf{x} \ {}_{s}Y_{\ell
  m}(\mathbf{x}) = s \sqrt{\frac{2\ell+1}{4\pi}}
\sqrt{\frac{(\ell-|s|)!}{(\ell+|s|)!}} \ A_\ell^s \ \delta_{m0} \ ,
\end{equation}
where the coefficients $A_\ell^s$ satisfy the relation $A_\ell^{-s} =
A_\ell^s$, and they are defined only for $\ell \ge |s|$. In
particular, in the case of $s=1,2$ we have
\begin{subequations}
\begin{equation}
A_\ell^1 = \begin{cases}
\frac{1}{4^{\ell+1}} \left( \begin{array}{c}
\ell+1 \\ \frac{\ell+1}{2}
\end{array} \right)^2 \frac{\ell+1}{\ell} \frac{\pi}{2} \ , & \ell \ \text{odd} \\
0 \ , & \ell \ \text{even}
\end{cases} \ ,
\end{equation}
\begin{equation}
A_\ell^2 = \begin{cases}
0 \ , & \ell \ \text{odd} \\
1 \ , & \ell \ \text{even}, \ \ell \neq 0
\end{cases} \ .
\end{equation}
\end{subequations}
The coefficients $A_\ell^1$ can be easily calculated by using the following
recurrence relation:
\begin{equation}
A_{\ell+2}^1 = \frac{\ell(\ell+2)}{(\ell+3)(\ell+1)} A_{\ell}^1 \ ,
\end{equation}
with initial values $A_{0}^1 = 0$ and $A_{1}^1 = 1/2$.

\acknowledgments
Partial financial support from the Spanish Ministerio de Econom\'{i}a
y Competitividad Projects AYA2012-39475-C02-01 and Consolider-Ingenio
2010 CSD2010-00064 is acknowledged.

\bibliographystyle{JHEP}
\bibliography{derivatives.bib}

\end{document}